%% file: DiffuDrift.tex
\documentclass[traditabstract]{aa} 

\usepackage{graphicx}
\usepackage{txfonts}
\usepackage{tikz}
\usepackage[colorlinks=true,urlcolor=dblue,citecolor=dblue,
            linkcolor=dblue]{hyperref}
\usepackage{ulem}
\usepackage{mhchem}

\def\Teff{T_{\rm eff}}
\def\nH{n_{\langle\rm H\rangle}}
\def\nHi{n_{\langle\rm H\rangle,i}}
\def\Vl{V_{\!\ell}}
\def\Nl{N_{\ell}}
\def\ek{\epsilon_k}
\def\pdiff#1#2{\frac{\partial #1}{\partial #2}}
\def\Dd{D_{\rm\hspace{0.05ex}d}}
\def\Sr{S_{\rm\hspace{-0.1ex}r}}
\def\nHleft{n_{\langle\rm H\rangle,1}}
\def\nHright{n_{\langle{\rm H}\rangle,I}}
\def\nHim1{n_{\langle\rm H\rangle,i-1}}
\def\nHip1{n_{\langle\rm H\rangle,i+1}}
\def\vdreq{{v^{\hspace{-0.8ex}^{\circ}}_{\rm dr}}}

\def\aap{A\&A}

\def\apjl{ApJL}
\def\apjs{ApJS}

\DeclareRobustCommand{\Eins}{\text{\usefont{U}{bbold}{m}{n}1}}

\def\vgas{\vec{\rm v}_{\rm gas}}

\def\Sr{S_{\!r}}
\def\vdrn{\vec{\rm v}_{\rm dr}^{\hspace{-0.9ex}^{\circ}}}
\def\er{\widehat{\vec{r}}}
\def\rhod{\rho_{\rm d}}
\def\St{{S\hspace*{-0.3ex}t}}
\def\taumix{\tau_{\rm mix}}

\def\la{\mathrel{\mathchoice {\vcenter{\offinterlineskip\halign{\hfil
$\displaystyle##$\hfil\cr<\cr\sim\cr}}}
{\vcenter{\offinterlineskip\halign{\hfil$\textstyle##$\hfil\cr
<\cr\sim\cr}}}
{\vcenter{\offinterlineskip\halign{\hfil$\scriptstyle##$\hfil\cr
<\cr\sim\cr}}}
{\vcenter{\offinterlineskip\halign{\hfil$\scriptscriptstyle##$\hfil\cr
<\cr\sim\cr}}}}}
\def\ga{\mathrel{\mathchoice {\vcenter{\offinterlineskip\halign{\hfil
$\displaystyle##$\hfil\cr>\cr\sim\cr}}}
{\vcenter{\offinterlineskip\halign{\hfil$\textstyle##$\hfil\cr
>\cr\sim\cr}}}
{\vcenter{\offinterlineskip\halign{\hfil$\scriptstyle##$\hfil\cr
>\cr\sim\cr}}}
{\vcenter{\offinterlineskip\halign{\hfil$\scriptscriptstyle##$\hfil\cr
>\cr\sim\cr}}}}}

\def\sll{\sffamily\itshape}

\begin{document} 

\input{farben.txt}

\title{Dust in brown dwarfs and extra-solar planets}

\subtitle{VII. Cloud formation in diffusive atmospheres}

\author{Peter Woitke\inst{1,2}
       \and
        Christiane Helling\inst{1,2,3}
       \and
        Ophelia Gunn\inst{4}
}

\institute{
           Centre for Exoplanet Science, University of St Andrews, 
           St Andrews, UK
         \and
           SUPA, School of Physics \& Astronomy, University of St
           Andrews, St Andrews,  KY16 9SS, UK 
         \and
           SRON Netherlands Institute for Space Research, Sorbonnelaan 2,
           3584 CA Utrecht, NL
         \and
           SUPA, School of Physics and Astronomy, University
           of Edinburgh, Edinburgh, EH9 3JZ, UK
}

\date{Received 10/07/2019; accepted 09/11/2019}

\abstract{The precipitation of cloud particles in brown dwarf and
  exoplanet atmospheres establishes an ongoing downward flux of
  condensable elements. To understand the efficiency of cloud
  formation, it is therefore crucial to identify and to quantify the
  replenishment mechanism that is able to compensate for these local
  losses of condensable elements in the upper atmosphere, and to keep
  the extrasolar weather cycle running. In this paper, we introduce a
  new cloud formation model by combining the cloud particle moment
  method of Helling \& Woitke with a diffusive mixing approach, taking
  into account turbulent mixing and gas-kinetic diffusion for both gas
  and cloud particles.  The equations are of diffusion-reaction type
  and are solved time-dependently for a prescribed 1D atmospheric
  structure, until the model has relaxed toward a time-independent
  solution.  In comparison to our previous models, the new hot Jupiter
  model results ($T_{\rm eff}\!\approx\!2000\,$K, $\log\,g\!=\!3$)
  show fewer but larger cloud particles which are more concentrated
  towards the cloud base. The abundances of condensable elements in
  the gas phase are featured by a steep decline above the cloud base,
  followed by a shallower, monotonous decrease towards a plateau, the
  level of which depends on temperature. The chemical composition of
  the cloud particles also differs significantly from our previous
  models. Due to the condensation of specific condensates like
  Mg$_2$SiO$_4$[s] in deeper layers, certain elements, such as Mg, are
  almost entirely removed from the gas phase early.  This leads to
  unusual (and non-solar) element ratios in higher atmospheric layers,
  which then favours the formation of SiO[s] and SiO$_2$[s], for example,
  rather than MgSiO$_3$[s]. Such condensates are not expected in
  phase-equilibrium models that start from solar abundances.  Above
  the main silicate cloud layer, which is enriched with iron and metal
  oxides, we find a second cloud layer made of Na$_2$S[s] particles in
  cooler models ($T_{\rm eff}\!\lessapprox\!1400\,$K).}




\maketitle

\section{Introduction}

The number of confirmed extrasolar planets has reached more than 4000,
but only a hand-full of them can be studied in detail (see
e.g. \citealt{2016ApJ...832..191N, 2017AJ....154...95H,
  2017AJ....153..138B, Arcangeli2018}). Indirect observations,
like transmission spectroscopy, have demonstrated the presence of
clouds (\citealt{2016Natur.529...59S, 2016ApJ...832..191N,
  Pino2018, 2017MNRAS.467.4591G, 2018MNRAS.474..876K,
  2018MNRAS.474.5485T}). Far easier targets for atmosphere studies are
brown dwarfs, which are very similar to planets with respect to their
physical parameters and atmospheric processes. The coolest brown dwarfs (Y
dwarfs) reach effective temperatures as low as 250\,K
(\citealt{2017ApJ...842..118L, 2014ApJ...786L..18L}). The observation
of brown dwarfs allows us to identify the vertical cloud structures
(\citealt{2013ApJ...768..121A, 2015ApJ...812..163B,
  2016ApJ...826....8Y, 2014A&ARv..22...80H}). To date, between 1500
and 2000 brown dwarfs are known (depending on whether late M-dwarfs and/or
early L-dwarfs are included; \citealt{2015ApJS..219...33G,
  2018ApJS..234....1B}) and are relatively well-studied compared to
the $\sim\!4000$ extrasolar planets, for which the era of spectral analysis
has only just begun.

Cloud formation has a profound impact on the remaining gas phase
abundances and radiative transfer effects, but cloud particles will
also affect the ionisation state of the atmosphere, which is well
known for solar system objects (\citealt{2016SGeo...37..705H,
  2016PPCF...58g4003H}). Efforts are therefore ongoing to construct
physical models describing the formation of clouds in exoplanet and
brown dwarf atmospheres. Such detailed models are necessary tools to
provide the context for observations and to uncover processes not
directly accessible by observations. Part of this effort is the
consistent coupling of cloud formation with 1D atmosphere models with
radiative transfer and convection (\citealt{Tsuji1996,2001ApJ...556..872A,
  Tsuji2002, Witte2009, 2012RSPTA.370.2765A, 2017A&A...608A..70J}; also
\citealt{2008MNRAS.391.1854H}), but also in 3D in order to study the
time-dependent climate of extrasolar planets
(\citealt{2016A&A...594A..48L,Lines2018a}) and to understand
observational implications beyond 1D
(\citealt{2017A&A...601A..22L,Lines2018a}).

As our understanding of cloud formation progresses
(e.g. \citealt{2015A&A...575A..11L,
  2017ApJ...847...89K, Hoerst2019}), including its
implication for habitability (\citealt{2015NatSR...513977N}), we start
to refine our approaches. One long-standing discussion is how to model
the element replenishment in 1D cloud forming atmospheres, because
without replenishment, a quasi-static atmosphere must be cloud free
(Appendix A in \citealt{Woitke2004}). \citet{2013A&A...558A..91P}
utilised passive tracers to study the atmospheric mixing in {3D
  (shallow water approximation)} simulations for irradiated, dynamic
but convectively stable atmospheres of (giant gas) planets. They
observe that cloud particles are distributed throughout the whole
atmosphere.

\citet{2013A&A...558A..91P} state: ``{\sl In statistical steady state,
  this upward dynamical flux balances the downward transport due to
  particle settling and allows the atmospheric tracer abundance to
  equilibrate at finite (non-zero) values despite the effect of
  particle settling. The mechanism does not require convection, and
  indeed, the vertical motions that cause the upward transport in our
  models are resolved, large-scale motions in the stably stratified
  atmosphere.  These vertical motions are a key aspect of the
  global-scale atmospheric circulation driven by the day-night heating
  contrast.}''  This assessment confirms our conclusion that the
upward transport of condensable elements through the atmosphere by
mixing is indeed the key to understand cloud formation. {However,
  challenges arise from the choice of the inner boundary condition
  (\citealt{2019arXiv190413334C}), chemical gradients
  (\citealt{2019ApJ...876..144T}), and the need to include cloud
  particle feedback in order to test mixing parameterisations. A
  particular interesting case will be the ultra-hot Jupiters where day
  and night-sides can be expected to have very distinct (vertical)
  mixing patterns and scales. In this paper, we consider self-luminous
  giant gas planets, for which the irradiation from their host stars
  is negligible, such as young giant gas planets and brown
  dwarfs. Brown dwarfs atmospheres are by now understood to be rather
  similar to giant gas planets, in particular atmospheres from
  low-gravity brown dwarfs and young gas giants
  (\citealt{2018ApJ...854..172C}).}

{\cite{2000Icar..143..244M} point out that large-scale mixing helps to
  homogenise a gravitationally stratified atmospheres consisting of
  different kinds of molecules.  This, however, only prevails up to a
  certain altitude {above which} gas-kinetic diffusion starts to
  dominate over mixing \citep{Zahnle2016}. Different approaches have
  been chosen to represent this vertical mixing in 1D atmosphere
  models (\citealt{2001ApJ...556..872A, Woitke2004,
    2008MNRAS.391.1854H, 2014IAUS..299..271A, 2017A&A...608A..70J})
  and in 3D models (\citealt{Lee2015,Lines2018}) by
  measuring vertical velocity fluctuations and deriving mixing
  parameterisations from 2D or 3D radiation-hydrodynamics simulations
  (\citealt{2002A&A...395...99L, 2010A&A...513A..19F,
    2013A&A...558A..91P,2018ApJ...866....1Z})}.

Cloud formation modelling becomes an increasingly important part also
of exoplanet/brown dwarf retrieval approaches for which, however,
computational speed is an essential limitation.  As part of the
ARCiS\footnote{ARtful modelling Code for exoplanet Science} retrieval
platform, \cite{2019A&A...622A.121O} presented a fast forward model
that consistently solves diffusive mixing and cloud particle growth
for exoplanet atmospheres.
  
In this paper, we present a new theoretical approach that consistently
combines cloud formation modelling with diffusive transport for
element replenishment. After presenting the main formula body of our
model in Sects.~\ref{ss:cfd} to \ref{ss:ec}, we summarise our ansatz
for handling the diffusion coefficient in Sect.~\ref{ss:DC}, before we
present our main results in Sects.~4 and 5.  We conclude in
Sect.~6. An overview of quantifying diffusion coefficients in the
literature is provided in Appendix~\ref{s:DG_A}.

\section{Cloud formation with diffusive transport of gas and cloud particles}

Cloud formation involves {at least} seed particle formation
(nucleation), surface growth and evaporation, element depletion,
gravitational settling and element replenishment. During their decent
through the atmosphere, cloud particles may change phase or, more
general, chemical composition, and may collide with each others
leading to further growth.  These cloud formation processes have been
described previously
\citep{Woitke2003,Helling2006,2013RSPTA.37110581H} and different cloud
formation models have been compared by \citet{2008MNRAS.391.1854H}
{with an update} by \citet{2018ApJ...854..172C}. We therefore only
provide a short summary here, a recent review can be found in
\citet{Helling2019b}.

Clouds are made of particles (aerosols, droplets, solid particles).
The formation of these particles requires condensation seeds, which
are produced, for the case of the Earth atmosphere, by volcano
eruptions, ocean sprays and wild fires.  In absence of these crucial
processes, which all require the existence of a solid planet surface,
cloud formation needs to start with the formation of seed particles
through chemical reactions in the gas phase, involving the formation
of molecular clusters. The formation of seed particles requires a
highly supersaturated gas. Once such seed particles become available,
many materials are already thermally stable {and can condense on
  these surfaces simultaneously}.  Nucleation and growth reduce the
local element abundances and have a strong feedback on the local
composition of the atmospheric gas. As macroscopic cloud particles
form, they display a spectrum of sizes as well as a mixture of
condensed materials.  The local particle size distribution and the
material mixture change as the cloud particles move through the
atmosphere (hence, encounter different thermodynamic conditions), for
example by gravitational settling (rain).  Particle-particle collision
will continue to shape the size distribution function. Cloud particles
may break up into smaller units (shattering) or stick together to form
even bigger units (coagulation). {Cloud particles may also be
  transported upward and downward by macroscopic mixing
  processes}. Particle-particle processes are not part of our present
model which focuses on the formation of cloud particles and their
feedback on the local chemistry through element
depletion/enrichment. We note that the surface growth does shut off
the nucleation process due to efficient element depletion
(\citealt{2015A&A...575A..11L}) such that a simultaneous treatment of
nucleation and growth is required in order to calculate the number of
cloud particles forming in the first place.

\subsection{Cloud formation as reaction-diffusion system}
\label{ss:cfd}

As introduced in \citet{Woitke2003}, we consider the evolution of the
size distribution function $f(V)\rm\,[cm^{-6}]$ of cloud particles in
the particle volume interval $V\,...\,V+dV$ as affected by advection,
settling, surface reactions and (new) by diffusion according to the
following master equation
\begin{equation}
  \frac{\partial\big(f(V)dV\big)}{\partial t} + 
  \nabla \Big(\vec{\rm v}(V)\,f(V)dV\Big) 
  ~=~ \sum_k R_k\,dV ~-~ \nabla \vec{\phi}_{\rm d}\,dV \ .
  \label{eq:master}
\end{equation}
$R_k$ are the various gain and loss rates due to surface chemical
reactions, which lead to growth and evaporation of the particles (see
Eqs.~59-62 for large Knudsen numbers, and Eqs.~68-71 for small Knudsen
numbers in \citealt{Woitke2003}).  The volume of the particles $V$ is
chosen as size variable to formulate the material deposit by surface
reactions in the most straightforward way.  The last term in
Eq.\,(\ref{eq:master}) accounts for the additional gains and losses
due to diffusive mixing. The cloud particle velocity $\vec{\rm v}(V)$
is assumed to be given by the hydrodynamical gas velocity $\vgas$ plus
a vertical equilibrium drift velocity $\vdrn(V)$
\begin{equation}
  \vec{\rm v}(V) = \vgas + \vdrn(V) \ .
\end{equation}
Applying Fick's first law \citep[see e.g.][]{Bringuier2013}, the
diffusive flux $\vec{\phi}_{\rm d}$ of the cloud particles in volume
interval $V\,...\,V+dV$ ($\vec{\phi}_{\rm d}\,dV$ has units
$\rm[cm^{-2}s^{-1}]$) is given by the concentration gradient of those
particles
\begin{equation}
  \vec{\phi}_{\rm d}\,dV = -\,\rho \Dd\nabla\left(\frac{f(V)dV}{\rho}\right), 
  \label{eq:jdust}
\end{equation}
where $\Dd\,\rm[cm^2s^{-1}]$ is the diffusion coefficient for those
cloud particles and $\rho\,\rm[g/cm^3]$ the gas
density. We introduce moments of the cloud particle size distribution
as
\begin{equation}
  \rho L_j = \int_{\Vl}^\infty\!\!f(V)V^{j/3}\,dV \ .
  \label{eq:Lj}
\end{equation}   
Multiplying Eq.\,(\ref{eq:master}) with $V^{j/3}$ and integrating over volume, 
we obtain the following system of moment equations for large Knudsen
numbers \citep[see details in][]{Woitke2003}
\begin{eqnarray}
  \frac{\partial (\rho L_j)}{\partial t} 
  &\hspace{-2.5mm}+\hspace{-2.5mm}& \nabla\big(\vgas\,\rho L_j\big)
  ~=~ \underbrace{\Vl^{j/3} J_\star}_{\rm nucleation}
  ~+~ \underbrace{\frac{j}{3}\,\chi\,\rho L_{j-1}}_{\rm growth}\nonumber\\
  &\hspace{-2.5mm}-\hspace{-2.5mm}& 
    \underbrace{\nabla\!\int_{V_l}^{\infty}\!\!\vec{\rm v}_{\rm dr}(V)
      f(V)\,V^{j/3}\,dV}_{\rm drift}  
   ~-~\underbrace{\nabla\!\int_{V_l}^{\infty}\!\!\vec{\phi}_{\rm
       d}\,V^{j/3}\,dV}_{\rm diffusion} \ ,
  \label{eq:mom_diff}
\end{eqnarray}
where $J_\star\rm\,[s^{-1}cm^{-3}]$ is the nucleation rate and
$\chi\rm\,[cm/s]$ the net growth velocity. For large Knudsen numbers
and subsonic velocities (Epstein regime), the equilibrium drift
velocity, also called final fall speed, is given by \citet{Schaaf1963}
\begin{equation}
  \vdrn = - \frac{\sqrt{\pi}\,g\,\rhod\,a}{2\,\rho\,c_T}\;\er \ ,
  \label{eq:vdr1}
\end{equation}
where $a$ is the particle radius, $\rhod$ the cloud particle
material density, $\er$ the unit vector pointing away from the centre
of gravity, and $g$ the gravitational
acceleration. $c_T\!=\!\sqrt{2kT/\bar{\mu}}$ is an abbreviation, $T$
the temperature, $k$ the Boltzmann constant, and $\bar{\mu}$ the mean
molecular weight of the gas particles.

Using Eq.\,(\ref{eq:vdr1}) with $a=\Big(\frac{3 V}{4\pi}\Big)^{1/3}$
and assuming that the particle diffusion coefficient $\Dd$ is
independent of size, we can carry out the integrations in
Eq.\,(\ref{eq:mom_diff}). The final result is
\begin{eqnarray}
  \pdiff{(\rho L_j)}{t} + \nabla(\vgas\,\rho L_j) 
    &\hspace{-2mm}=\hspace{-2mm}& \Vl^{j/3} J_\star
    ~+~\frac{j}{3}\,\chi\,\rho L_{j-1}  \nonumber\\
     &\hspace{-2mm}+\hspace{-2mm}&
       \nabla\!\left(\xi\,\frac{\rhod}{c_T}\,L_{j+1}\;\er\right)
    ~+~\nabla\left(\Dd\,\rho\,\nabla L_j\right) 
  \label{eq:LwithDiff}
\end{eqnarray}
with abbreviation 
\begin{equation}
  \xi = \frac{\sqrt{\pi}}{2}\left(\frac{3}{4\pi}\right)^{\!1/3} \!g
  \ .
  \label{eq:xi}
\end{equation}
A size-dependent diffusion coefficient, $\Dd(V)$, would lead to an open set of
moment equations as discussed by \citet{2013RSPTA.37110581H}.

\begin{table}
\caption{Variable definitions and units}
\label{tab:defs}
\vspace*{-2mm}
\resizebox{90mm}{!}
{\begin{tabular}{l|l|l}
\hline
symbol  & description & unit\\
\hline
$z$                       & vertical coordinate     & cm\\
$\nH$                     & hydrogen nuclei density & cm$^{-3}$\\
$\rho = \mu_{\rm H}\,\nH$  & gas mass density        & g\,cm$^{-3}$\\
$T$                       & gas temperature         & K\\
$a$                       & cloud particle radius   & cm\\
$V=\frac{4\pi}{3}a^3$     & volume of a cloud particle & cm$^3$\\
$V^{\rm s}$                & volume occupied by material s & cm$^3$ \\  
$\Vl$                     & minimum volume of a cloud particle & cm$^3$\\
$f(V)$                    & size distribution function & $\rm cm^{-6}$\\
$\rho L_j$                & cloud particle moments           & cm$^{j-3}$\\
$L_j$                     & $j^{\rm th}$ moment           & cm$^j$\,g$^{-1}$\\
$J_\star$                  & nucleation rate        & $\rm cm^{-3}s^{-1}$\\
$\chi$                    & net growth speed       & cm/s\\
$\rho_{\rm d}$             & mean cloud particle material density & $\rm g\,cm^{-3}$ \\
$\phi$                    & diffusive flux         & $\rm cm^{-2}\,s^{-1}$\\
$\Dd$                     & cloud particle diffusion coefficient  & cm$^2$\,s$^{-1}$\\
$D_{\rm gas}$              & gas diffusion coefficient  & cm$^2$\,s$^{-1}$\\
$s$                       & index for different solid materials  & $1\,...\,S$\\
$r$                       &  index for the surface reactions & $1\,...\,R$\\
\hline  
\end{tabular}}
\end{table}


\subsection{Generalisation to mixed materials}

We assume that all cloud particles are perfect spheres with well-mixed
material composition which is independent of size, but depends on time
and location in the atmosphere \citep{Helling2006,Helling2008}. Using
the index $s=1\,...\,S$ to distinguish between the different solid
materials, for example Al$_2$O$_3$[s], TiO$_2$[s], Mg$_2$SiO$_4$[s]
and Fe[s], we write
\begin{equation}
  V \!=\! \sum\limits_s V^s            \;\mbox{,}\,\;
  L_3 \!=\! \sum\limits_s L_3^s        \;\mbox{,}\,\;
  J_\star \!=\! \sum\limits_s J_\star^s \;\mbox{,}\,\;
  \chi \!=\! \sum\limits_s\,\chi^s     \;\mbox{,}\,\;
  b^s \!=\! \frac{L_3^s}{L_3}          \,,\hspace*{-3mm}
\end{equation}
where $b^s$ is the volume fraction of material $s$ in the cloud
particles\footnote{In \citet{Helling2008}, we have used the
  notation $b^s=V_s/V_{\rm tot}$ where $V_s=\rho
  L_3^s=\int_{V_{\ell}}^\infty f(V) V^s\,dV$ is the volume occupied by
  solid $s$ per volume of stellar atmosphere and $V_{\rm tot}=\rho
  L_3=\int_{V_{\ell}}^\infty f(V) V\,dV$ is the total solid volume per
  volume of stellar atmosphere.}. The mean cloud particle material
density is given by $\rhod=\sum_s b^s\rho_s$ where $\rho_s$ is the
mass density of a pure material $s$. Most materials will not nucleate
themselves ($J_\star^s=0$), but will use alien nuclei to grow
on. Using this approximation, we can split the third moment equation
into a set of third moment equations for single materials as follows
\begin{eqnarray}
  \pdiff{(\rho L_3^s)}{t} + \nabla(\vgas\,\rho L_3^s) 
  &\!\!=\!\!& \Vl J_\star^s 
          ~+~ \chi^s \rho L_2 \nonumber\\
  &\!\!+\!\!& \nabla\!\left(\xi\,\frac{\rhod}{c_T}\,b^s L_4\,\widehat{r}\,\right)
  ~+ \nabla\left(\Dd\,\rho\nabla L_3^s\right) \ .
  \label{eq:L3mat}
\end{eqnarray}
Adding up Eqs.\,(\ref{eq:L3mat}) for all solids $s$ again yields
Eq.\,(\ref{eq:LwithDiff}) for $j\!=\!3$. The different materials grow
at different speeds which depend on the amount of available atoms and
molecules in the gas phase and on the supersaturation ratio. Islands of some
materials may grow whereas others are thermally unstable and shrink.
This behaviour is obtained by summing up the contributions of all
surface reactions $r=1\,...\,R$ \citep[for examples see
  Table~1 in][]{Helling2008}
\begin{equation}
  \chi = \sum\limits_s\,\chi^s = \sum\limits_s \sum\limits_r c_r^s V_0^s \ ,
\end{equation}
where $V_0^s\rm\,[cm^3]$ is the volume of one unit of material of kind
$s$ in the solid state and $c_r^s\rm\,[cm^{-2}\,s^{-1}]$ is the
effective surface reaction rate
\begin{equation}
  c_r^s ~=~ \sqrt[3]{36\pi}\;
      \frac{\nu_r^s\,n_r^{\rm key}{\rm v}_r^{\rm rel}\,\alpha_r}{\nu_r^{\rm key}}
      \left(1-\frac{1}{\Sr}\right) \times \left\{\begin{array}{ll}
                                         1 & \mbox{if\ \ } \Sr\geq 1 \\
                                       b^s & \mbox{if\ \ } \Sr<1
                                   \end{array}\right. \ ,
  \label{eq:cr}
\end{equation}
where $n_r^{\rm key}$ is the particle density [cm$^{-3}$] of the key
species of surface reaction $r$, $\alpha_r$ the sticking probability,
$\nu_r^{\rm key}$ its stoichiometric factor in that reaction, ${\rm
  v}_r^{\rm rel}=\sqrt{kT/(2\pi\,m^{\rm key})}$ its thermal relative
velocity and $m^{\rm key}$ its mass. These growth rates are derived
from a simple hit-and-stick model where we usually assume
$\alpha_r\!=\!1$. The impact of the limited number of known
$\alpha_r\!\not=\!1$ has been studied by \citet{Helling2006}. $\Sr$ is
  the reaction supersaturation ratio as introduced in \citep[][see
    their App.~B]{Helling2006}.  For example, in the reaction
\begin{equation}
 \rm 2\,CaH ~+~ 2\,Ti ~+~ 6\,H_2O ~\longleftrightarrow~ 2\,CaTiO_3[s] ~+~ 7\,H_2
\end{equation}
the key species is either CaH or Ti, depending on which species is
less abundant. $\nu_r^{\rm key}\!=\!2$ in either case,
$s\!=\rm\!CaTiO_3$[s] is the solid species, and $\nu_r^s\!=\!2$ units
of CaTiO$_3$[s] are produced by one reaction.

We note that Eq.\,(\ref{eq:cr}) differs slightly from our previous
definition \citep[Eq.\,4 in][]{Helling2008}. The new
growth/evaporation rates now always change sign at $\Sr\!=\!1$ as they
should, independent of the value of $b^s$. When supersaturated
($\Sr\!>\!1$), we assume that the total surface of the particles acts
as a funnel to collect the impinging molecules from the gas phase,
followed by fast hopping to find a place on a matching island of kind
$s$ (see Fig.~1 in \citealt{Helling2006}). But for under-saturation
($\Sr\!<\!1$), we assume that the molecules triggering the evaporation
processes must hit one of the islands of the matching kind, the
probability of which is $b^s$.

\subsection{Element conservation with diffusive replenishment}
\label{ss:ec}

An integral part of our cloud formation model is the element
conservation.  We must identify a replenishment mechanism which is
able to compensate for the losses of elements due to cloud particle
formation and settling in the upper atmosphere.  In this paper, we
include diffusion of gas particles along their concentration gradients.
As cloud particles form, they consume certain elements in the upper
atmosphere, creating a negative element abundance gradient. Thus, gas
particles containing those elements will ascent diffusively in that
atmosphere.  Analogous to the formulation of the master equation for
the dust particles, we formulate the element conservation as
diffusion-reaction system
\begin{eqnarray}
  \pdiff{(\nH\ek)}{t} + \nabla(\vgas\,\nH\ek)
  &\!\!\!=\!\!\!& -\sum\limits_s \nu_k^s \Nl J_\star^s 
    ~-~ \rho L_2 \sum\limits_s\sum\limits_r \nu_k^s c_r^s  \nonumber\\ 
 & &+~ \nabla\left(D_{\rm gas}\,\nH\nabla \ek\right)  \ ,
\label{eq:epsDiff}
\end{eqnarray}
{where $\epsilon_k$ is the abundance of element $k$ with respect
to hydrogen.}      
The chemical reactions leading to nucleation and growth appear as
negative source terms here, because they consume elements.  We choose
$\nH$ as density variable in Eq.\,(\ref{eq:epsDiff}), {the total
  hydrogen nuclei particle density, which is proportional to $\rho$ in
  hydrogen-dominated atmospheres}. $\nH\ek$ is the total number
density [cm$^{-3}$] of nuclei of element $k$ in any chemical
form. $\nu_k^s$ is the stoichiometric factor of element $k$ in solid
$s$, for example $\nu_{\rm Ti}^{\rm TiO2[s]}\!=\!1$, and $D_{\rm gas}$
[cm$^2$ s$^{-1}$] is the gas diffusion coefficient. For simplicity, we
assume that all molecules are transported by the same diffusion
coefficient, which is valid within a factor 2 or 3 for gas-kinetic
diffusion (sometimes called the binary diffusion coefficient, see
Eq.\,(\ref{eq:defD}) and App.\,\ref{s:DG_A}), and is entirely justified
when eddy diffusion dominates. The involved diffusive gas element flux
$\vec{\phi}_k^{\,\rm diff}$ [cm$^{-2}$s$^{-1}$] is given by
\begin{equation}
  \vec{\phi}_k^{\,\rm diff} = -\,D_{\rm gas}\,\nH\,\nabla \ek \ .
  \label{eq:jelm}
\end{equation}

\subsection{Gas diffusion coefficient}
\label{ss:DC}
The diffusion coefficients provide the kinetic information to calculate
the transport rates from concentration gradients
\citep[e.g.][]{Lamb2011}. In general, gas and cloud particles diffuse
with different efficiencies because of their different inertia and
collisional cross sections with the surrounding gas.

\begin{figure}
   \hspace*{-2mm}\includegraphics[width=94mm]{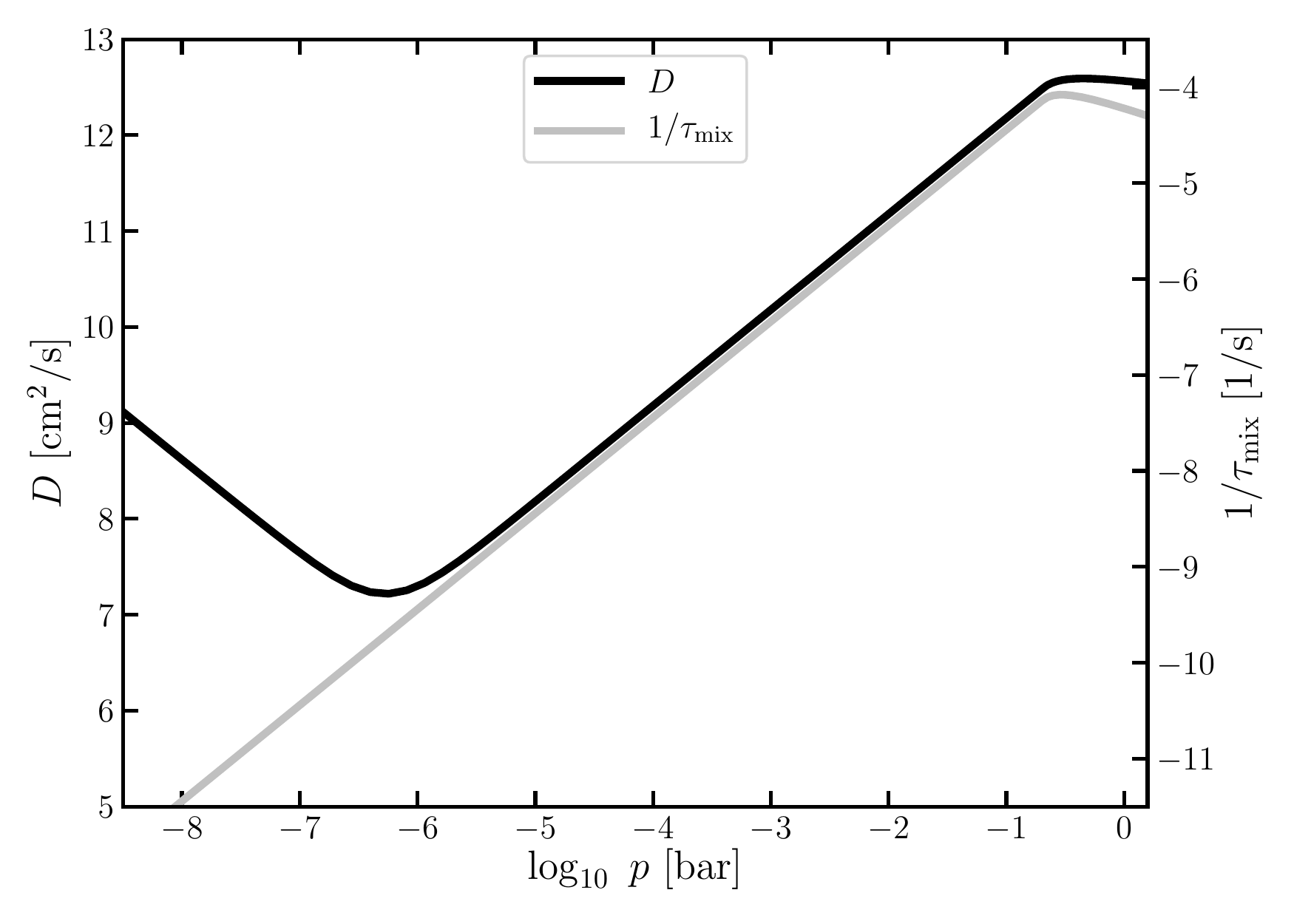}
   \vspace*{-7mm}
   \caption{The gas diffusion coefficient $D_{\rm gas}$
     (Eq.~\ref{eq:Dgas}) in the new {\sc DiffuDrift} model for a brown
     dwarf atmosphere model with $T_{\rm eff}\!=\!1800\,$K, $\log
     g\!=\!3\,\rm[cm/s^2]$ and solar abundances.  The grey line shows
     the inverse mixing timescale $\tau_{\rm mix}$ as assumed in the
     previous {\sc Drift} model.  $\tau_{\rm mix}$ is calculated
     according to Eq.\,(9) in \citep{Woitke2004}. Both quantities are
     computed for $\beta\!=\!\beta^{\prime}\!=\!1$ and both $y$-axes
     show exactly 8 orders of magnitude.}
   \label{fig:Dgas}
   \vspace*{-1mm}
\end{figure}

The determination of the gas diffusion coefficient $D_{\rm gas}$ is of
crucial importance for our model. We include gas-kinetic diffusion and
large-scale turbulent (eddy) diffusion as mixing processes. The
gas-kinetic diffusion coefficient is given by
\begin{equation}
  D_{\rm micro} = \frac{1}{3}\,{\rm v_{th}}\,\ell \ ,
  \label{eq:defD}
\end{equation}
where $\ell=1/(\sigma\,n)$ is the mean free path, $n$ the total gas
particle density and $\sigma\approx 2.1\times 10^{-15}\rm\,cm^2$ a
typical cross-section for collisions between the molecules under
consideration with H$_2$. The thermal velocity is defined as ${\rm
  v}_{\rm th}=\sqrt{8 k T/(\pi\,m_{\rm red})}$ where $m_{\rm red}$ is
the reduced mass for collisions between the molecule and H$_2$
\citep{Woitke2003}.  This gas-kinetic diffusion $\propto\!1/n$ is
negligible in the lower high-density layers of brown dwarf and
planetary atmospheres, where instead mixing by large-scale turbulent
or convective motions is the dominant mixing process
\citep{2001ApJ...556..872A, Ludwig2002, Woitke2004,
  2012RSPTA.370.2765A, 2013A&A...558A..91P, Lee2015}. The large-scale
(turbulent/convective/eddy) gas diffusion coefficient is given by
\begin{equation}
  D_{\rm mix} \approx \langle {\rm v}_z\rangle\,L
  \quad\quad\mbox{with}\quad\quad
  L = \alpha\,H_p\ ,
  \label{eq:defDmix}
\end{equation} 
where $\langle {\rm v}_z\rangle$ is the root-mean-square average of
the fluctuating part of the vertical velocity in the atmosphere,
considering averages over sufficiently large volumes and/or long
integration times, $L$ is the mixing length, and $H_p$ is the local
pressure scale height. {$\alpha$ is a dimensionless parameter of the
  order of one.  We use $\alpha\!=\!1$ in this work, but note that
  $\alpha$ can be fine-tuned to describe the actual mixing scales as
  revealed by detailed hydrodynamic modelling.} Inside the convective
part of the atmosphere $\langle {\rm v}_z\rangle\approx {\rm v}_{\rm
  conv}$ is assumed, where ${\rm v}_{\rm conv}$ is the convective
velocity, which is an integral part of stellar atmosphere models,
derived from mixing length theory in 1D models.  Above the convective
atmosphere, where the Schwarzschild criterion for convection is false,
$\langle {\rm v}_z\rangle$ decreases rapidly with increasing $z$, but
never quite reaches zero due to convective overshoot \citep[see
  e.g.][]{2016ApJ...832....6B}. We apply a power-law in $\log p$ to
approximate this behaviour
\begin{equation}
  \log\,\langle {\rm v}_z\rangle ~=~ \log {\rm v}_{\rm conv} 
  - {\beta^{\prime}}\cdot\max\Big\{0,\log p_{\rm conv}-\log p(z)\Big\}
  \label{eq:convm}
\end{equation}
with a free parameter ${\beta^{\prime}}\approx 0.0\,...\,2.2$
\citep{Ludwig2002}. The total gas diffusion coefficient is then
\begin{equation}
  D_{\rm gas} ~=~ D_{\rm mix} + D_{\rm micro} \ .
  \label{eq:Dgas}
\end{equation}
At high altitudes, the gas density $n$ is small and hence $D_{\rm
  micro}$ is large, whereas $D_{\rm mix}$ is small when
${\beta^{\prime}}\!>\!0$. Therefore, at some pressure level in the
atmosphere, the gas-kinetic diffusion will start to dominate.
Figure~\ref{fig:Dgas} shows a typical structure as assumed in our
models. The minimum of $D_{\rm gas}$ around $10^{-3}$\,mbar
corresponds to the crossover point (called the {\it homopause}), upward of
which $D_{\rm micro}$ dominates and the atmospheres is not
well-mixed. \citet{2000Icar..143..244M} draw similar conclusions
concerning Saturn's atmosphere. The maximum of $D_{\rm gas}$ around
$p_{\rm conv}\!=\!0.2$\,bar results from the start of the convective
layer, within which both $\langle {\rm v}_z\rangle$ and $D_{\rm gas}$
are approximately constant. {Appendix~\ref{s:DG_A} summarises
some of the formulas currently applied in the literature for the
gas diffusion coefficient.}

\subsection{Cloud particle diffusion coefficient}

{The diffusion of solid particles due to turbulent gas
  fluctuations was studied, in consideration of protoplanetary discs,
  by \citet{Dubrulle1995}, \citet{2004ApJ...614..960S},
  \citet{Youdin2007} and \citet{Riols2018}. All works apply mean field
  theory (also called Reynolds decomposition ansatz), where the
  densities and velocities of both the particles and the gas are
  decomposed into a mean component (that depends only on $z$) and a
  small fluctuating part.

The response of the solid particles to the turbulent gas variations is
then determined by comparing two timescales.  The stopping or
frictional coupling timescale is given by
\begin{equation}
  \tau_{\rm stop} = \frac{a\,\rhod}{v_{\rm th}\,\rho}
  \quad\quad\mbox{with}\quad\quad
  v_{\rm th} = \sqrt{\frac{8\,kT}{\pi\,\bar{\mu}}}\ ,
  \label{eq:tstop}
\end{equation}
where $a$ the particle radius. Equation~(\ref{eq:tstop}) follows from
a general relaxation ansatz $\tau_{\rm stop}\!=\!m\,\big(\partial
F_{\rm fric}/\partial v_{\rm dr}\big|_{\,\vdreq}\big)^{-1}$, see
Eq.\,(21) in \citet{Woitke2003}, for the special case of large Knudsen
numbers in a subsonic flow (the so-called Epstein regime), which we
assume is valid here.

The second timescale is the eddy turnover or turbulence correlation
timescale $\tau_{\rm eddy}(l)$ in consideration of a spectrum of
different turbulent modes associated with different wave-numbers $k$ or
different spatial eddy sizes $l$.  In a Kolmogorov type of power
spectrum $P(k)\!\propto\!k^{-5/3}$, any given cloud particle of size $a$
tends to co-move with all sufficiently large and slow turbulent eddies
whereas its inertia prevents following the short-term, small
turbulence modes.


In order to arrive at an effective particle diffusion coefficient, the
advective effect of all individual turbulent eddies has to be
averaged, and thereby transformed into a collective diffusive effect.
This procedure is carried out with different procedures and
approximations. The result of \citet[][see their
  Eq.\,27]{2004ApJ...614..960S}, reads
\begin{equation}
  \Dd = \frac{D_{\rm mix}}{1+\St}
  \quad\quad\mbox{with}\quad\quad
  \St=\frac{\tau_{\rm stop}}{\tau_{\rm eddy}} \ .
  \label{eq:Stokes}
\end{equation}
where $\Dd$ is the size-dependent cloud particle diffusion coefficient
and $\St$ is the Stokes number of the particle in 
consideration of the largest eddy size $L$. The eddy turnover
timescale of the largest turbulence mode is given by
\begin{equation}
  \tau_{\rm eddy} = \frac{L}{\langle {\rm v}_z\rangle} \ .
\end{equation}
Both the size of the largest eddy $L$ and the average of the
fluctuating part of the vertical velocity $\langle {\rm v}_z\rangle$
are assumed to be identical to the mixing length and velocity
appearing in Eq.\,(\ref{eq:defDmix}). Combining the above
equations we find}
\begin{equation}
  \St = \frac{a\,\rhod\,D_{\rm mix}}{v_{\rm th}\,\rho\;L^2} \ .
\end{equation}
The impact of the size dependence of $\Dd$ on the cloud particle
moments was explored by \citet{2013RSPTA.37110581H}, who showed that
this leads to an open set of moment equations, which seems impractical
for an actual solution. In the frame of this work, we will therefore
only explore the two limiting cases of very large and very small
Stokes numbers throughout the atmosphere. For small particles with
$\St\!\ll\!1$, we have $\Dd\!\to\!D_{\rm mix}$ and for huge particles
$\St\!\to\!\infty$, we have $\Dd\!\to\!0$.
\begin{equation}
  \hspace*{1mm}
  \begin{array}{lll}
    \mbox{case 1:} &
    \Dd = D_{\rm mix} & \mbox{if all cloud particles are small,}\\
    \mbox{case 2:} &
    \Dd = 0          & \mbox{if all cloud particles are large.}
  \end{array}
\end{equation}
Our results show that both approximations lead to rather similar cloud
structures in the models explored so far, i.e.\ the inclusion of
turbulent cloud particle motions does not seem to be a critical
ingredient to our present model. {However, in preliminary models for
hot Jupiters, where $D_{\rm mix}(z)$ is more flat or even increasing
with height, this might be different.}

\section{Static plane-parallel atmosphere}

Before we proceed with the numerical solution of the full time-dependent
model of cloud formation in diffusive media, we first discuss the 1D static
case in order to better understand the expected results from these equations.
Considering the plane-parallel ($\nabla\to d/dz$), static
($\vgas\!=\!0$) and stationary case ($\partial/\partial t\!=\!0$), our
Eqs.~(\ref{eq:LwithDiff}), (\ref{eq:L3mat}) and (\ref{eq:epsDiff})
simplify to
\begin{eqnarray}
 0\!\!&=&\!\! \Vl^{j/3} J_\star
    + \frac{j}{3}\,\chi\,\rho L_{j-1}
    + \xi\frac{d}{dz}\!\left(\frac{\rhod}{c_T}L_{j+1}\right) 
    + \frac{d}{dz}\!\left(\Dd\,\rho\frac{d L_j}{dz}\right) 
 \label{eq:stat1D1}\\
 0\!\!&=&\!\!\Vl J_\star^s 
    + \rho L_2\!\sum\limits_r c_r^s V_0^s 
    + \xi\frac{d}{dz}\!\left(\frac{\rhod}{c_T}\,b^s L_4\right)
    + \frac{d}{dz}\!\left(\Dd\,\rho \frac{dL_3^s}{dz}\right) 
 \label{eq:stat1D2}\\
 0\!\!&=&\!\!-\sum\limits_s \nu_k^s \Nl J_\star^s 
    - \rho L_2 \sum\limits_s\sum\limits_r \nu_k^s c_r^s 
    + \frac{d}{dz}\!\left(D_{\rm gas}\,\nH\frac{d\ek}{dz}\right)  \ .
    \label{eq:stat1D3}
\end{eqnarray}

\subsection{The total element fluxes}
\label{sec:tindep}

In the hydrostatic stationary case, the total
vertical flux of elements (due to vertical settling of cloud particles
{\it and\,} diffusive transport) must be zero everywhere in the
atmosphere and for each element. This conclusion can be derived
formally by adding together Eq.\,(\ref{eq:stat1D3}) and $\sum_s$
(Eq.~\ref{eq:stat1D2})\,$\cdot\,\nu_k^s/V_0^s$, using
$V_\ell\!=\!N_\ell V_0^s$. The chemical source terms (nucleation and
growth terms) cancel out exactly, and in case $\Dd\!=0$ we find
\begin{eqnarray}
  \frac{d}{dz}\!\left(D_{\rm gas}\,\nH\frac{d\ek}{dz}\right) 
  ~+~ \xi\frac{d}{dz}\!\left(
      \frac{\rhod}{c_T}L_4\sum\limits_s\frac{\nu_k^s\,b^s}{V_0^s}\right)
  ~=~0  \nonumber\\
  \quad\Rightarrow\quad
  \underbrace{D_{\rm gas}\,\nH\frac{d\ek}{dz}}_{\displaystyle-\phi_k^{\rm\,diff}} ~+~
  \underbrace{\xi\,\frac{\rhod}{c_T}L_4\sum\limits_s\frac{\nu_k^s\,b^s}{V_0^s}
             }_{\displaystyle \phi_k^{\rm\,settle}}
  ~=~ {\rm const}_{\,k}
  \label{statepsk}
\end{eqnarray}
$\phi_k^{\rm\,diff}$ [cm$^{-2}$ s$^{1}$] is the upward element flux by diffusion
in the gas phase and $\phi_k^{\rm\,settle}$ is the downward flux of
elements contained in the settling cloud particles at this
point. Equation (\ref{statepsk}) would still allow for solutions with
constant (i.e.\ time-independent and height-independent) total element
fluxes throughout the atmosphere, but this would require matching
feeding and removing rates at the bottom and the base of the
atmosphere, which does not seem to be physically plausible.
Thus, ${\rm const}_{\,k}\!=\!0$ and we find
\begin{equation}
  \phi_k^{\rm\,diff} = \phi_k^{\rm\,settle}
  \;\;\;\Rightarrow\;\;\;
  \frac{d\ek}{dz} ~=~
  -\,\frac{\xi\,\rhod L_4}{c_T\,D_{\rm gas}\,\nH} \sum\limits_s\frac{\nu_k^s\,b^s}{V_0^s}
  ~\leq~0 \ .
  \label{eq4}
\end{equation}
According to Eq.~(\ref{eq4}), the element abundance gradients in
cloudy, static ($\vec{\rm v}_{\rm gas}\!=\!0$) and stationary
($\partial/\partial t\!=\!0$) atmospheres must be negative because of
the downward transport of elements via the precipitation of cloud
particles, which must be balanced by an upward directed diffusive flux
of elements in the gas phase, which requires a negative concentration
gradient.  This conclusion is correct whenever cloud particles are
present $(L_4\!>\!0)$ and gravity is active $(\xi\!>\!0)$, otherwise
the gas element abundances are constant.  The abundance gradients of
different elements are proportional to the element composition of the
settling cloud particles at this point. The abundances of all elements
$k$ involved in cloud formation \,{\it must decrease monotonically}\,
toward the top of the atmosphere.

\section{Numerical solution of the time-dependent cloud formation problem}

Equations (\ref{eq:stat1D1})\,$-$\,(\ref{eq:stat1D3}) form a system of
$(3\!+\!S\!+\!K)$ coupled 2$^{\rm nd}$ order differential equations, which can
be transformed into a system of $2\times(3\!+\!S\!+\!K)$ 1$^{\rm
  st}$ order ordinary differential equations (ODEs). Unfortunately, we have not
been able to solve this ODE system directly. The boundary conditions
are partly given at the lower and partly at the upper boundary of the
model, see Sect.~\ref{sec:bc}. The integration direction must be
downward in order to model the nucleation of new cloud
particles. Hence, we tried a shooting method where $\ek(z_{\rm max})$
is varied at the top of the atmosphere until $\ek(z_{\rm min})$ is met,
i.e.\ the given values in the deep atmosphere. We found it impossible to
proceed this way. A tiny change of $\ek(z_{\rm max})$ in the 12$^{\rm
  th}$ digit was still observed to change $\ek(z_{\rm min})$ by a
factor of two. The reason for this extreme sensitivity seems
to be the nucleation rate with its threshold character as function of
supersaturation, and hence as function of $\ek$.

Looking for alternatives, we found that a simulation of the
time-dependent equations on a given vertical grid can be performed by
means of the operator splitting method as explained in
Sect.~\ref{sec:os}. We evolve the atmospheric cloud structure
$\big\{L_j(z,t), L_3^s(z,t), \ek(z,t)\big\}$ for a sufficiently long
time, until it approaches the time-independent case
$\big\{L_j^\circ(z),L_3^{s,\circ}(z),\ek^\circ(z)\big\}$, which is the
stationary structure we are interested in.  Assuming a plane-parallel
($\nabla\to d/dz$) and static ($\vgas\!=\!0$) atmosphere,
Eqs.~(\ref{eq:LwithDiff}), (\ref{eq:L3mat}) and (\ref{eq:epsDiff})
read, including the time-dependent terms
\begin{eqnarray}
 \frac{d}{dt}\Big(\rho L_j\Big) &\!\!\!=\!\!\!& \Vl^{j/3} J_\star
    ~+~ \frac{j}{3}\,\chi\,\rho L_{j-1} \label{eq:statD1}\\
    &\!\!\!+\!\!\!& \xi\frac{d}{dz}\!\left(\frac{\rhod}{c_T}L_{j+1}\right) 
    ~+~ \frac{d}{dz}\!\left(\Dd\,\rho\frac{d L_j}{dz}\right) 
    \hspace{5mm}(j=1,2,3)
 \nonumber\\
 \frac{d}{dt}(\rho L_3^s) &\!\!\!=\!\!\!& \Vl J_\star^s 
    ~+~ \chi^s \rho L_2 \label{eq:statD2}\\
    &\!\!\!+\!\!\!& \xi\frac{d}{dz}\!\left(\frac{\rhod}{c_T}\,b^s L_4\right)
    ~+~ \frac{d}{dz}\!\left(\Dd\,\rho \frac{dL_3^s}{dz}\right) 
    \hspace{4mm}(s=1\,...\,S)
 \nonumber\\
 \frac{d}{dt}(\nH\ek) &\!\!\!=\!\!\!& -\sum\limits_s \nu_k^s \Nl J_\star^s 
    ~-~ \rho L_2 \sum\limits_s\sum\limits_r \nu_k^s c_r^s \label{eq:statD3}\\
    &\!\!\!+\!\!\!& \frac{d}{dz}\!\left(D_{\rm gas}\,\nH\frac{d\ek}{dz}\right) 
    \hspace{24mm}(k=1\,...\,K)  \ .
    \nonumber
\end{eqnarray}

\subsection{Closure condition}
The moment Eqs.~(\ref{eq:statD1}) and (\ref{eq:statD2}) are not closed
because $L_4$ appears twice of the right side, a consequence of larger
particles settling faster (Eq.\,\ref{eq:vdr1}). Therefore, a numerical
solution requires a closure condition as
\begin{equation}
  L_4 = {\cal F}(L_0,L_1,L_2,L_3)   \ .
\end{equation}
We use the closure condition explained in the appendix~A.1 of
\citep{Helling2008}. The idea is to approximate the particle size
distribution $f$ by a double $\delta$-function which has four
parameters. These parameters are determined by matching the given
moments $L_0$, $L_1$, $L_2$ and $L_3$, and result in the forth moment
$L_4$ according to the definition of the dust moments
(Eq.\,\ref{eq:Lj}).

\subsection{Operator splitting method}
\label{sec:os}

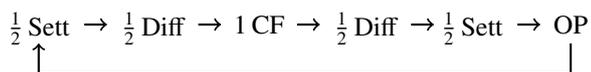
\begin{figure}
\centering
    \begin{tikzpicture}
        \node[] at (0.0,1) {$\frac{1}{2}$\,Sett};
        \draw[black, thick,->] (0.6,1.05) -> (0.9,1.05);
        \node[] at (1.5,1) {$\frac{1}{2}$\,Diff};
        \draw[black, thick,->] (2.1,1.05) -> (2.4,1.05);
        \node[] at (2.9,1.05) {1\,CF};
        \draw[black, thick,->] (3.4,1.05) -> (3.7,1.05);
        \node[] at (4.3,1) {$\frac{1}{2}$\,Diff};
        \draw[black, thick,->] (4.9,1.05) -> (5.2,1.05);
        \node[] at (5.7,1) {$\frac{1}{2}$\,Sett};
        \draw[black, thick,->] (6.3,1.05) -> (6.6,1.05);
        \node[] at (7.0,1.05) {OP};
        \draw[black, thick,-] (7.0,.8) -> (7.0,0.4);
        \draw[black, thick,-] (7.0,.4) -> (0,0.4);
        \draw[black, thick,->]  (0,.4) -> (0,0.8);
    \end{tikzpicture}
\caption{Operator splitting calling sequence. Sett\,$=$\,gravitational
  settling, Diff\,$=$\,diffusion, CF\,$=$\,cloud formation
  (nucleation, growth and evaporation), and OP\,$=$\,output.
  1/2 means half a timestep and 1 a full timestep.}
\label{fig:os}
\vspace*{-2mm}
\end{figure}

Figure \ref{fig:os} visualises our numerical approach using the
operator splitting method \citep{Klein1995}.
\begin{enumerate}
\setlength{\itemsep}{2pt}
\setlength{\parskip}{0pt}
\setlength{\topsep}{0pt}
\setlength{\parsep}{0pt}
\setlength{\partopsep}{0pt}
\item We update $L_j$ and $L_3^s$ only according to the settling
  source terms (the terms on the right side of Eqs.\,(\ref{eq:statD1})
  and (\ref{eq:statD2}) containing $L_{j+1}$ and $L_4$), applying half
  a timestep $\Delta t/2$, see details in App.~\ref{app:sett}.
\item We call the diffusion solver for half a timestep $\Delta t/2$ to
  update $\ek$ and, optionally, $L_j$ and $L_3^s$, if the cloud
  particles are to be diffused as well, see App.~\ref{app:diff}.
\item We integrate the chemical source terms (nucleation, growth and
  evaporation) for a full timestep $\Delta t$. These equations are
  stiff at high densities and require an implicit integration scheme.
  We use the implicit ODE-solver {\sc Limex 4.2A1}
  \citep{Deuflhard1987}.  The computation of the chemical source terms
  on the r.h.s.\ proceeds as follows: (i) for given temperature $T$,
  density $\nH$ and element abundances $\ek$, we call the equilibrium
  chemistry code {\sc GGchem} \citep{Woitke2018} to calculate all
  molecular concentrations; (ii) those results are used to calculate
  the reaction supersaturation ratios $\Sr$; (iii) the nucleation
  rates $J_\star^s$ and net surface reaction rates $c_r^s$ are
  determined.
\item We finish the timestep by calling again the diffusion solver
  for $\Delta t/2$ and the settling solver for $\Delta t/2$ in this order.
\item Checkpoint and output files are written for visualisation.
\end{enumerate}
The method is computationally limited by the time consumption for the
implicit integration of the chemical source terms, which requires
numerous calls of the equilibrium chemistry. This is why we do not
split CF (Fig.~\ref{fig:os}) but put it with a full timestep in the
centre of the operator splitting calling sequence. The cloud formation
part of the code is parallelised and can be executed for all
atmospheric layers independent of each other.

\subsection{Timestep control}
In order to produce accurate 2$^{\rm nd}$ order solutions, the
timestep must be limited to make sure that each operator remains in
the linear regime. For example, the sole application of a
cloud-chemistry timestep must not change the amount of dust or the
element abundances substantially in any computational cell. In order
to achieve code stability and accuracy, we limit the timestep as
follows:
\begin{enumerate}
\setlength{\itemsep}{2pt}
\setlength{\parskip}{0pt}
\setlength{\topsep}{0pt}
\setlength{\parsep}{0pt}
\setlength{\partopsep}{0pt}
\item The cloud particles must not jump over layers by settling
  \begin{equation}
  \Delta t ~<~ 0.5 \frac{\Delta z}{{\rm v}_{{\rm dr},j}}
  \end{equation}
  where $\Delta z$ is the vertical grid resolution and ${\rm v}_{{\rm
      dr},j}$ is the mean drift velocity affecting moment $\rho L_j$ as
  given by Eq.\,(\ref{eq:vj}). {This is the usual
  Courant-Friedrichs-Lewy (CFL) criterion to stabilise explicit
  advection scheme, with an additional safety-factor 1/2.}
\item Nucleation and cloud particle growth and evaporation, as
  integrated over $\Delta t$, must not change any of the gas element
  abundances by more than a given maximum relative change (default
  accuracy is $15\%$).
\item The timestep must not exceed the maximum explicit diffusion
  timestep (Eq.\,\ref{tstep}). 
\end{enumerate}
If one of these criteria becomes false during the simulation, the timestep is 
discarded and $\Delta t$ reduced. If, on the contrary, the criteria
are met easily, $\Delta t$ is increased for the subsequent timestep.

\subsection{Boundary conditions}
\label{sec:bc}
As our upper boundary condition, we assume that there are no cloud
particles settling into the model volume from above ${\rm v}_{{\rm
    dr},j}(z_{\rm max})\!=\!0$.  In the diffusion solver, we use a
zero-flux (closed box) upper boundary condition, i.e.\ the gradients of
$\ek$ are assumed to be zero at $z\!=\!z_{\rm max}$. The same applies
to the cloud particle moments $\frac{d}{dz}L_j(z_{\rm max})=0$ if they
are to be diffused as well.

The lower boundary is placed well below the main silicate cloud layer
to make sure that the temperature is too high to allow for any 
cloud particles to exist near the lower boundary $L_j(z_{\rm
  min})\!=\!0$. We also demand that the element abundances at the
lower boundary equal the given values as present deep in the
atmosphere $\ek(z_{\rm min}) = \ek^0$, where the $\ek^0$ are
considered as free parameters, for example solar abundances
\citep{Asplund2009}.

\subsection{Initial conditions}
\label{sec:ic}
We start all simulations from a cloud-free atmosphere
$L_j(z,t\!=\!0)=0$.  Concerning the element abundances, we have
experimented with two cases: (1) an `empty' atmosphere
$\ek(z,t\!=\!0)\!=\!0$ or (2) a `full' atmosphere
$\ek(z,t\!=\!0)\!=\!\ek^0$, where the index $k$ is applied to all
elements which can potentially be transformed completely into solids
($k\!=$\,Si, Mg, Fe, Al, Ti, ...), but not H, He, C, N, O, etc.
For the latter elements we put $\ek(z,t\!=\!0)\!=\!\ek^0$ in both
cases.  We found an identical final structure in both cases (see
App.~\ref{app:tests}), which is very reassuring. The models calculated
from initial condition (2), however, need much more computational time
to complete. In this case, the nucleation rate is initially huge and a
very large number of tiny cloud particles are created shortly after
initialisation, which take a long time to settle down in the
atmosphere.

In order to reach the final relaxed, time-independent state, the model
must be evolved until (i) the atmosphere is completely replenished
several times by fresh elements ascending diffusively from the lower
boundary to the very top and (ii) new grains formed high in the
atmosphere have sufficient time to settle down to the cloud base and
evaporate.  In comparison, the chemical processes are typically quite
fast.  We need to evolve one model for about $10^6$ timesteps, which,
depending on global parameters like $T_{\rm eff}$ and $\log\,g$,
translates into real evolutionary times 
between a few months to a few tens of years. On a parallel cluster,
one can complete one such model in a few days real time when using 16
processors (about 500 CPU\,hours), where the chemical equilibrium
solver {\sc GGchem} is called a few $10^9$ times.

\section{Results}

\subsection{Comparison to our previous cloud formation model}

In the previous Helling\,\&\,Woitke cloud formation models
\citep{Woitke2004,Helling2006,Helling2008}, henceforth called the {\sc
  Drift} models, the replenishment of elements was treated in a
different way, using a prescribed mass exchange timescale $\taumix(z)$
to replenish the atmosphere with fresh elements from the deep as
$\nH(\ek^0-\ek)/\taumix$. The mass exchange timescale was approximated
by a powerlaw $\log\taumix(z)\!=\!{\rm const}-\beta\log p(z)$ with
power-law index $\beta\!=\!2.2$ to describe convectional overshoot, see
equation 9 in \citet{Woitke2004}.  This simple approach led to an
ODE-system which can be solved within about 2\,CPU-min.

\begin{figure}
  \hspace*{-3mm}
  \includegraphics[width=92mm,trim=15 15 10 5,clip]{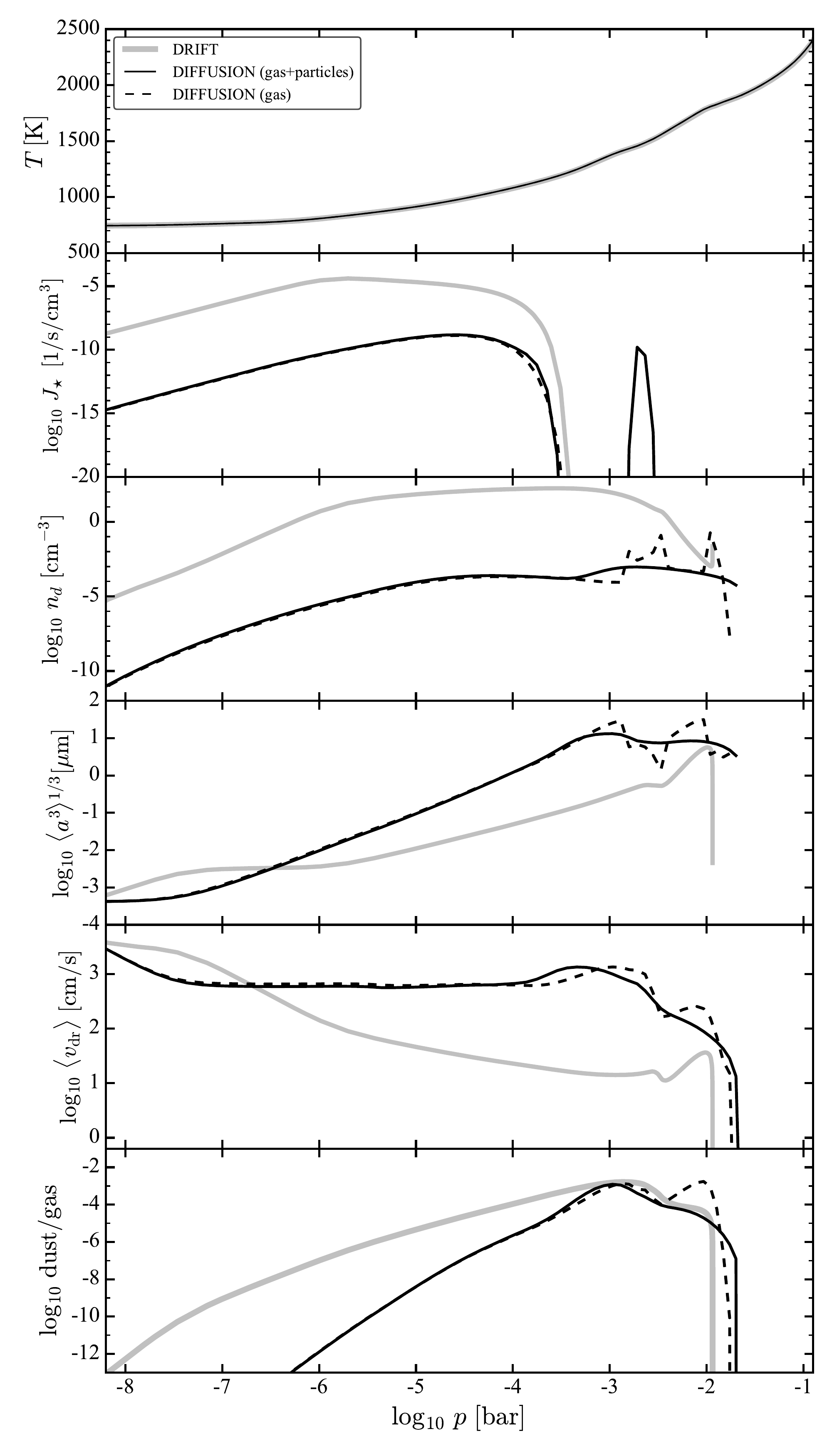}
  \vspace*{-5mm}
  \caption{Comparison of cloud formation models for $T_{\rm
      eff}\!=\!1800\,$K, $\log\,g\!=\!3$, metallicity $Z\!=\!1$, and
    $\beta\!=\!\beta^{\prime}\!=\!1$. The previous {\sc Drift} model is
    shown by the thick grey lines.  Two {\sc DiffuDrift} models are overplotted
    assuming pure gas diffusion (dashed lines) and gas $\!+\!$
    particle diffusion (black solid lines). dust/gas\,$=\!\rhod L_3$
    is the dust-to-gas mass ratio, $n_{\rm d}\!=\!\rho L_0$ the number
    density of cloud particles, $\langle
    a^3\rangle^{1/3}\!=\!\big(3\,L_3/(4\pi\,L_0)\big)^{1/3}$ the
    mass-mean particle radius, and $\langle{\rm v}_{\rm
      dr}\rangle\!=\!  \langle
    a^3\rangle^{1/3}\sqrt{\pi}\,g\,\rhod\,/\,(2\,\rho\,c_T)$ the
    corresponding drift velocity according to Eq.\,(\ref{eq:vdr1}).}
  \label{fig:oldnew}
  \vspace*{-1mm}
\end{figure}

Figure~\ref{fig:oldnew} compares the results of a previous {\sc Drift}
model with the new diffusive model, henceforth called the {\sc
  DiffuDrift} model. Both approaches model seed formation, kinetic surface
growth/evaporation of cloud particles and gravitational settling in the same
way, but differ in the treatment of the mixing which enters the cloud
formation and the element conservation equations. The underlying
temperature/pressure structures for all models discussed in this paper
are taken from a the {\sc Drift-Phoenix} atmosphere grid
\citep{2007IAUS..239..227D, hell2008, Witte2009,
  2011A&A...529A..44W}. In Fig.~\ref{fig:oldnew} we have selected a
model with effective temperature $T_{\rm eff}\!=\!1800\,$K, surface
gravity $\log\,g\!=\!3$ and metallicity $Z\!=\!1$ (i.e.\ solar
abundances are assumed deep in the atmosphere). The {\sc
  Drift-Phoenix} models solve the complete 1D model atmosphere problem
including convection, radiative transfer and hydrostatic structure,
coupled to our previous {\sc Drift} model, where the cloud
opacities are calculated by Mie and effective medium theory. The
resulting atmospheric structure are frozen for this study, i.e.\ the
feedback of the new cloud formation model on the ($p,T$)-structure is
not included.

The chemical setup for this comparison has 16 elements (H, He, Li, C,
N, O, Na, Mg, Si, Fe, Al, Ti, S, Cl, K, Ca), one nucleation species
(TiO$_2$), 12 solid species (TiO$_2$[s], Al$_2$O$_3$[s], CaTiO$_3$[s],
Mg$_2$SiO$_4$[s], MgSiO$_3$[s], SiO[s], SiO$_2$[s], Fe[s], FeO[s],
MgO[s], FeS[s], Fe$_2$O$_3$[s]) and 60 surface reactions. The
molecular setup in the new models is not quite identical, since the
{\sc Drift} model uses a previous version of the chemical equilibrium
solver {\sc GGchem}, which has been replaced by the latest version
\citep{Woitke2018} in the {\sc DiffuDrift} model. We use 189 molecules
in {\sc Drift} and 308 molecules in {\sc DiffuDrift} to find the
molecular concentrations in chemical equilibrium. {We do not see
  any substantial differences in molecular concentrations caused by this data
  update, unless the local temperature falls below about 400\,K.} Also
the thermochemical data for the selected solids is not entirely
identical, {but these differences are not substantial either, because
the local temperatures remain above 700\,K
in this test}.  We assume the mixing powerlaw index to be
$\beta=\beta'\!=\!1$ for both $\taumix(z)$ in {\sc Drift} and $D_{\rm
  gas}(z)$ in {\sc DiffuDrift}, see Eq.\,(\ref{eq:convm}) and
Fig.\,\ref{fig:Dgas}, {albeit the meaning of $\beta$ and
  $\beta^{\prime}$ is slightly different. We note that using
  $\beta'>\beta$ would likely produce results that are more similar to
  each other than those presented in this paper.} The lower volume
boundary for the size integration of the cloud particle moments is set
to $V_\ell\!=\!10 \times V_{\rm TiO_2}$ where $V_{\rm
  TiO_2}\!=\!3.14\times 10^{-23}\rm\,cm^{3}$ is the assumed volume of
one unit of solid TiO$_2$[s].


\begin{table}[!b]
  \vspace*{-2mm}
  \caption{Comparison of cloud column densities [mg/cm$^2$]$^\star$
    for the three models shown in Fig.~\ref{fig:oldnew} and discussed
    in the text.}
  \label{tab:oldnew}
  \vspace*{-2mm}
  \centering
  \def\z{\hspace*{-2mm}}
  \resizebox{80mm}{!}{\begin{tabular}{c|ccc}
  \hline
 {condensate} & {\sc Drift} & \z{\sc DiffuDrift}\z & \z{\sc DiffuDrift}\z \\
              &          & \z{dust\,$+$\,gas diffusion}\z & gas diffusion\z \\
  \hline
  &&&\\[-2.2ex]
  \ce{TiO2}     & $6.8\times 10^{-3}$ & $4.3\times 10^{-3}$ & $9.7\times 10^{-3}$ \\
  \ce{Al2O3}    & $0.57$              & $0.47$             & $7.9$ \\ 
\z\ce{MgSiO3}\z & $0.29$              & $0.040$            & $0.092$ \\ 
\z\ce{Mg2SiO4}\z& $0.54$              & $0.59$             & $0.71$ \\ 
  \ce{SiO}      & $0.48$              & $0.034$            & $0.038$ \\ 
  \ce{SiO2}     & $0.14$              & $6.0\times 10^{-3}$ & $9.1\times 10^{-3}$ \\ 
  \ce{Fe}       & $1.7$               & $0.70$             & $0.97$ \\ 
  \ce{FeO}      & $1.2\times 10^{-3}$ & $2.5\times 10^{-6}$ & $3.8\times 10^{-6}$ \\ 
  \ce{MgO}      & $0.3$               & $6.9\times 10^{-7}$ & $1.8\times 10^{-6}$ \\ 
  \ce{FeS}      & $2.4\times 10^{-3}$ & $5.5\times 10^{-5}$ & $1.0\times 10^{-4}$ \\ 
\z\ce{CaTiO3}\z & $0.040$             & $0.022$            & $0.32$ \\ 
\z\ce{Fe2O3}\z  & $7.3\times 10^{-7}$ & $4.2\times 10^{-11}$& \z$9.2\times 10^{-11}$\z\\
  \hline
  &&&\\[-2.2ex]
  total         & 4.0                 & 1.9                & 10.1 \\  
  \hline
  \end{tabular}}\\[1mm]
  \footnotesize{$^\star$: column densities are calculated as
  ~$\Sigma^s\!=\!\int\!\rho_s\,\rho L_3^s\,dz$\,.}
  \vspace*{-1mm}
\end{table}

The resulting cloud structures, as predicted by our previous {\sc
  Drift} and the new {\sc DiffuDrift} models, are compared in
Fig.~\ref{fig:oldnew}.  The diffusive transport of condensable
elements up into the high atmosphere with {\sc DiffuDrift} is much
less efficient than compared to the assumed replenishment in the {\sc
  Drift} model. As these elements are slowly mixed upwards by
diffusion, they can collide with existing cloud particles to condense
on, and hence much less of these elements reach the high atmosphere
where the nucleation takes place.  This is the main difference between
the {\sc Drift} and the {\sc DiffuDrift} models. In the previous {\sc
  Drift} models, the mixing process was assumed to take place
instantly.

\medskip\noindent
{\sll Cloud structure:\ } Consequently, the new {\sc DiffuDrift} model
is featured by up to 5 orders of magnitude lower nucleation rates
(Fig.\ref{fig:oldnew}) and less cloud particles high in the
atmosphere. At intermediate pressures ($10^{-6}...\,10^{-3}$\,bar) we
find that the fewer cloud particles in the {\sc DiffuDrift} model
grow quickly and reach particle sizes of about 10\,$\mu$m at 1\,mbar,
wheres in the {\sc Drift} model, since there are so many of them, the
cloud particles remain smaller, about 0.3\,$\mu$m. The growth of the
cloud particles is limited by the amount of condensable elements
available per particle, and therefore, this effect is expected.

The dust-to-gas mass ratio, $\rho_{\rm d}/\rho_{\rm gas}$, increases
more steeply in the {\sc DiffuDrift} model, but reaches about the same
maximum of order $10^{-3}$ at 1\,mbar as in the {\sc Drift}
model. Thus, overall, the cloud formation process is about equally
effective, but the clouds are spatially more confined in the {\sc
  DiffuDrift} model, reaching up just a few scale heights above the
cloud base.

Table~\ref{tab:oldnew} lists vertically integrated cloud column
densities for the three models. We find values of a few milli-grams of
condensates per cm$^2$, where the {\sc DiffuDrift} model without cloud
particle diffusion is found to be the most dusty one. Using an order
of magnitude estimate of cloud particle opacities (see
Appendix~\ref{sec:opac}), {values between several 100\,cm$^2$/g to
  several 1000\,cm$^2$/g at $\lambda\!=\!1\mu$m are expected,
  depending on material and particle size distribution}, i.e.\ a
column density of 1\,milli-gram of condensate per cm$^2$ roughly
corresponds to an optical depth of one at $\lambda\!=\!1\mu$m.  This
implies that all three models discussed here have optically thick
cloud layers.

The computation of more realistic cloud particle opacities will need
to take into account the height-dependent material composition, size
and possibly shape distribution of the cloud particles, as done, for
example, by \citet{2007IAUS..239..227D}, \citet{Witte2009,
  2011A&A...529A..44W} and \citet{Helling2019}. {In comparison to
  the previous {\sc Drift} models, the particles in the {\sc
    DiffuDrift} models are larger, which is likely to cause the
  optical depths to be somewhat smaller, although the cloud mass column
  densities are similar.} In addition, molecular opacities need to be
added to calculate the spectral appearance of the objects and feedback
onto the $(p,T)$-structure, which goes beyond the scope of the present
paper.

The resulting particle properties in the main cloud layer below
1\,mbar depend not only on the treatment of mixing, but also whether
or not we switch on the dust diffusion in the {\sc DiffuDrift}
model.  In this region, the cloud particles stepwise purify chemically
as they decent in the atmosphere (see Fig.~\ref{fig:VolComp}).
Thermodynamically less stable materials like Fe$_2$O$_3$[s], FeO[s]
and FeS[s] sublimate off the cloud particles sooner. Subsequently, the
abundant magnesium-silicates MgSiO$_3$[s] and Mg$_2$SiO4[s] sublimate
as well, which causes the cloud particles to shrink significantly
around 1\,mbar. Close to the cloud base, at about 1800\,K in this
model, only the most refractory materials remain, in particular metal
oxides such as Al$_2$O$_3$[s], CaTiO$_3$[s] and TiO$_2$[s], before
even these materials eventually sublimate and the cloud particles
evaporate completely.

As one material sublimates, the liberated elements may re-condense
into different condensates, which are thermodynamically more stable,
leading to rapid changes in particle size and material composition.
There is also a dynamical effect. When the particles shrink, their
fall speeds decrease which leads to spatial accumulation, hence the
number density of cloud particles $n_d$ increases. While these
effects and the general behaviour of the cloud particles are similar
in all three models, the steps of sublimation are more pronounced in
the {\sc DiffuDrift} model without dust diffusion. Dust diffusion
tends to smooth out the variations of particle size and density.

\begin{figure}
  \includegraphics[width=90mm,trim=16 41 10 11,clip]{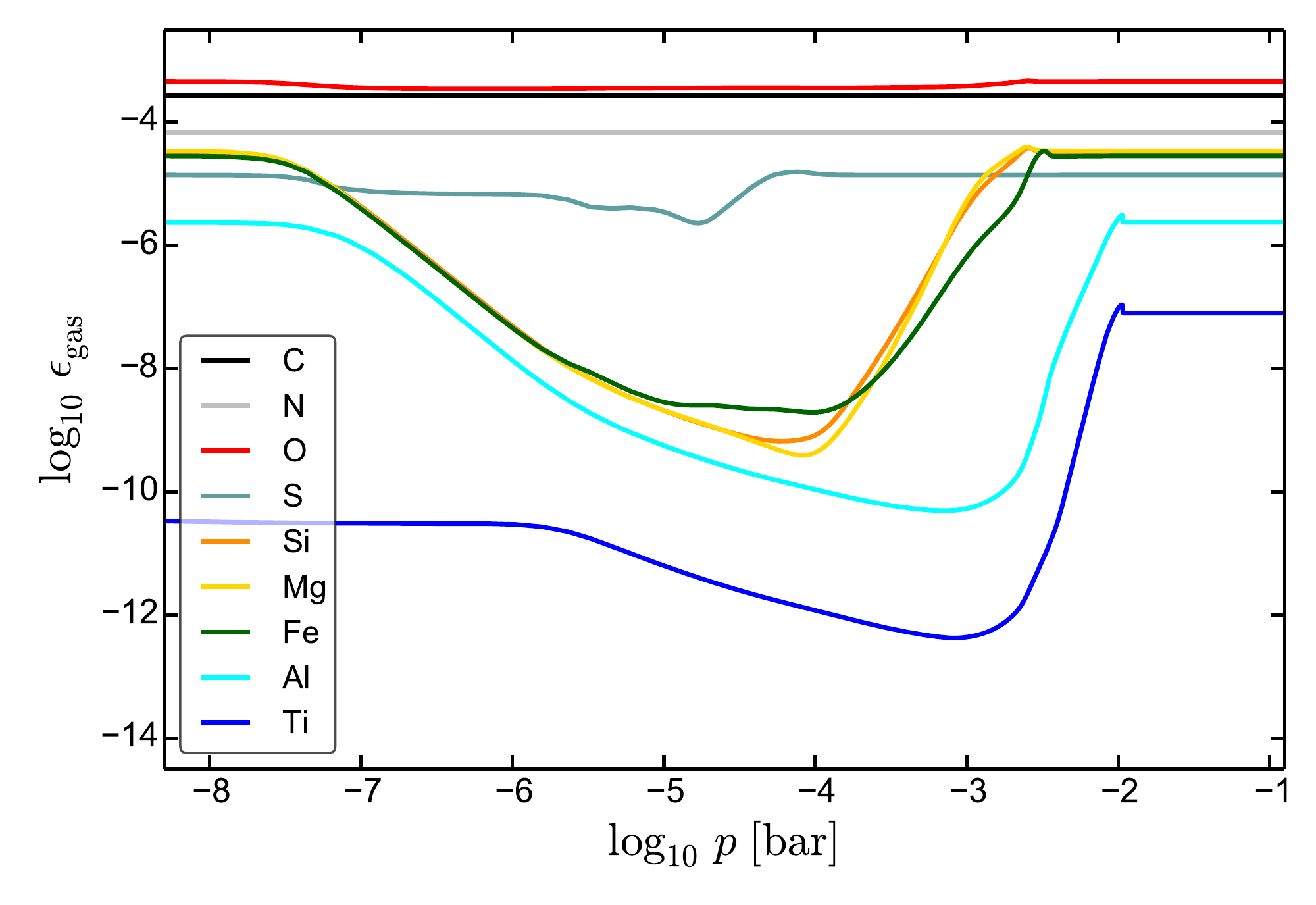}\\[-3mm]
  \includegraphics[width=90mm,trim=16  0 10 11,clip]{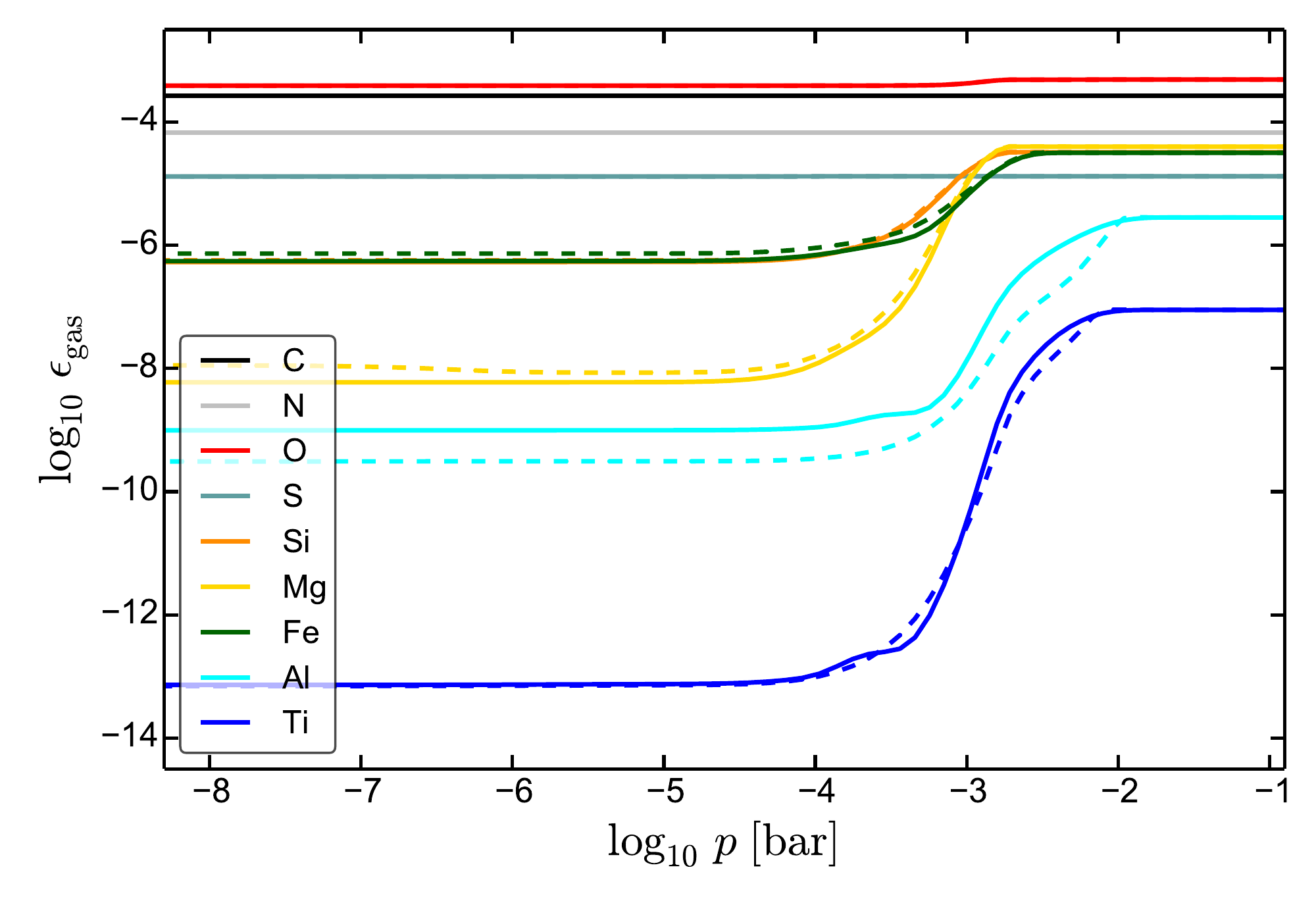}\\[-7mm]
  \caption{The impact of our assumptions about the mixing processes in
    the atmosphere on the resulting gas element abundances, in models
    with the same parameters as in Fig.~\ref{fig:oldnew}.  The {\bf
      upper plot} shows $\epsilon_k^{\rm gas}$ for instantaneous
    mixing as assumed in the previous {\sc Drift} model. The {\bf lower
      plot} shows the results according to the new {\sc DiffuDrift}
    models, where the full lines show the results for gas and dust
    diffusion, and the dashed lines show the results for gas diffusion
    only.}
  \label{fig:oldnew_eps}
  \vspace*{-2mm}
\end{figure}

\medskip\noindent {\sll Element abundances:\ }
Figure~\ref{fig:oldnew_eps} compares the resulting gas element
abundances. We see a strong depletion of condensable elements in the
main cloud layer in all three models, by up to 5 orders of magnitude,
concerning elements Ti, Al, Mg, Si, Mg and Fe.  However, the details
are different. The previous {\sc Drift} model is featured by minimums
of $\epsilon_k$ that are similar in depth as compared to the overall
decrease of $\epsilon_k$ in the {\sc DiffuDrift} models. High up in
the atmosphere, where cloud particles are virtually absent, there is
no surface to condense on, and so the instantaneous mixing assumption
in the {\sc Drift} models causes a re-increase of $\epsilon_k$ toward
the top of the atmosphere, unless the element can form nuclei.
In the extremely low density gas at these heights, these
nuclei simply fall through the atmosphere without much interaction,
whereas elements, which cannot nucleate, accumulate.

In contrast, in the new {\sc DiffuDrift} models, the abundances of all
elements involved in cloud formation decrease with height in a
monotonic way. This behaviour is expected in the final,
time-independent, relaxed state of the atmosphere as discussed in
Sect.~\ref{sec:tindep}.  In the stationary case, the downward
transport of condensable elements via the falling cloud particles must
be compensated by an upward diffusive transport of these elements in
the surrounding gas, which implies negative element abundance
gradients throughout the cloudy atmosphere, see Eq.\,(\ref{eq4}).  The total
drop of element abundances is deepest for Ti, but less deep for Si and
Fe as compared to Mg.

\cite{2010A&A...513A..19F} performed two-dimensional radiation
hydrodynamical simulations of substellar atmospheres which included a
time-dependent description for the formation of a single kind of cloud
particles for a fixed concentration of seed particles. The paper
discusses substellar objects with $T_{\rm
  eff}\!=\rm\!900\,K\!-\!2800\,K$, $\log (g)\!=\!5$ and solar element
abundances. Their Fig.~9 (bottom left panel) shows the fraction of
condensable gas in the atmosphere as a function of pressure, very
similar to our Fig.~\ref{fig:oldnew_eps} (lower plot).  These results
of Freytag et al.\ support our new {\sc DiffuDrift} results, where
abundances of condensable elements in the gas phase are decreasing
fast in the cloud layers, and stay about constant above the clouds.
We note that Freytag et al.\ have prescribed the number of seed
particles and considered only one generic condensate in their
simulations.

\begin{figure}
  \centering
  \includegraphics[width=90mm,trim=10 52 0 12,clip]{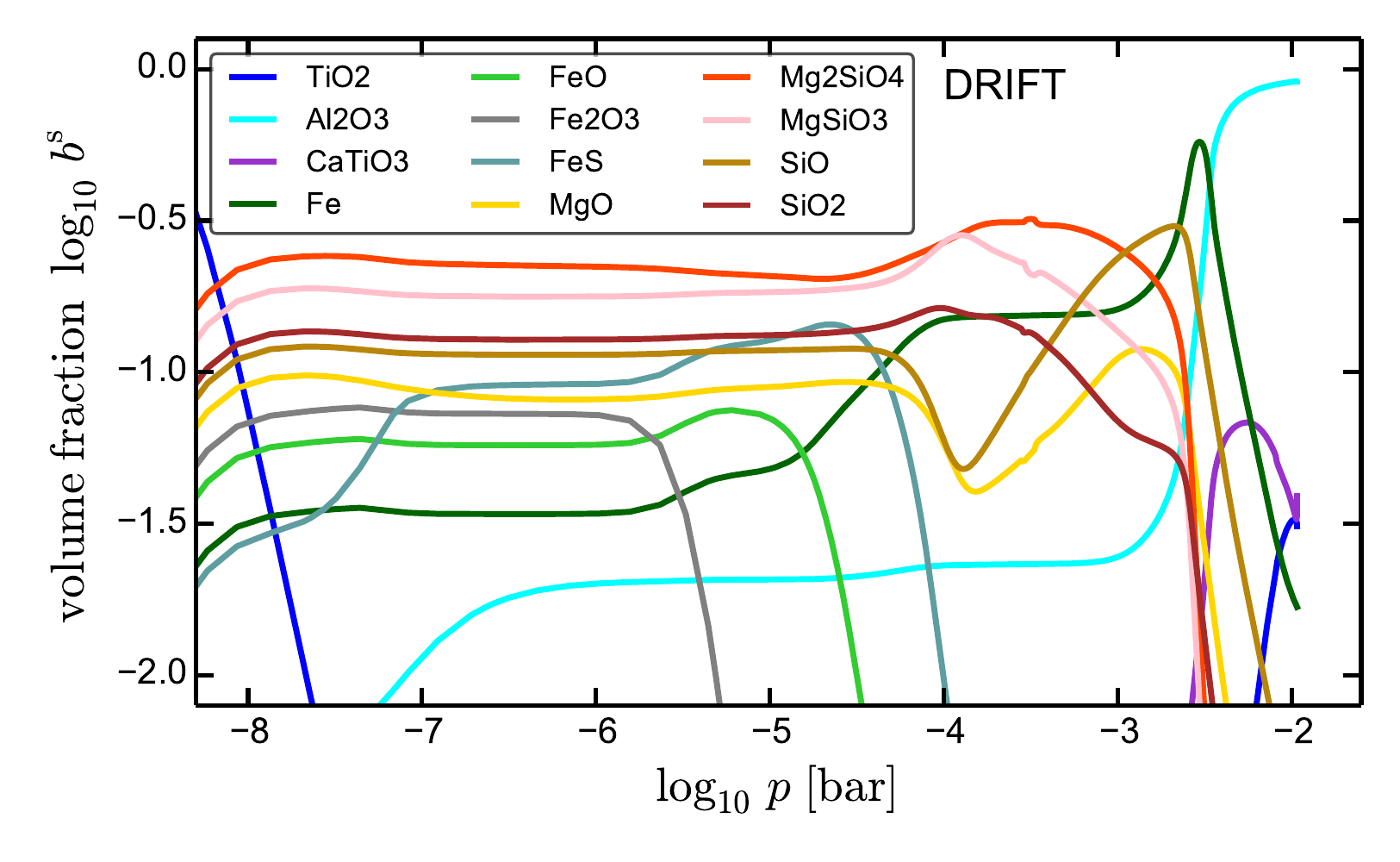}\\
  \includegraphics[width=90mm,trim=10 52 0 12,clip]{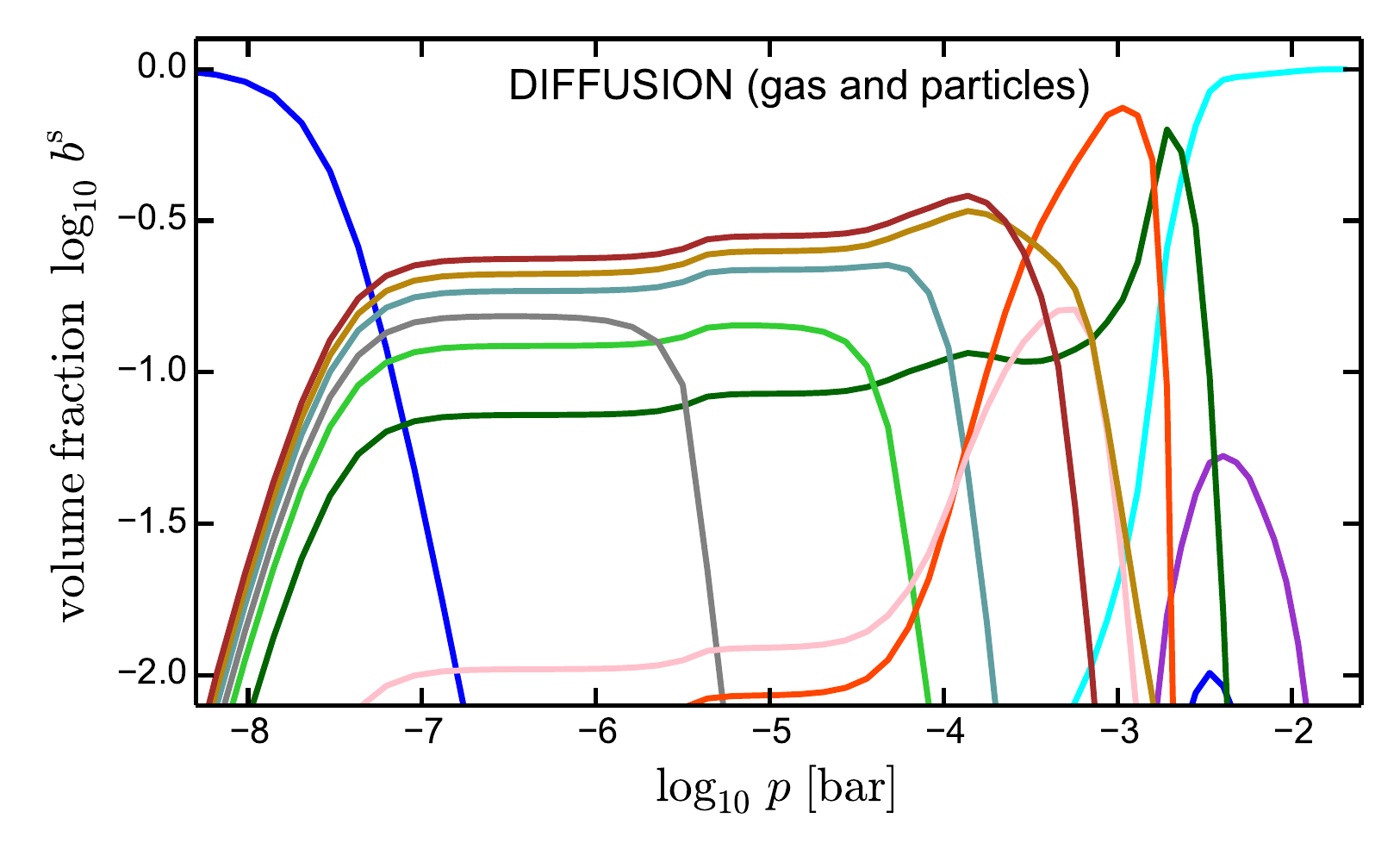}\\
  \includegraphics[width=90mm,trim=10 15 0 12,clip]{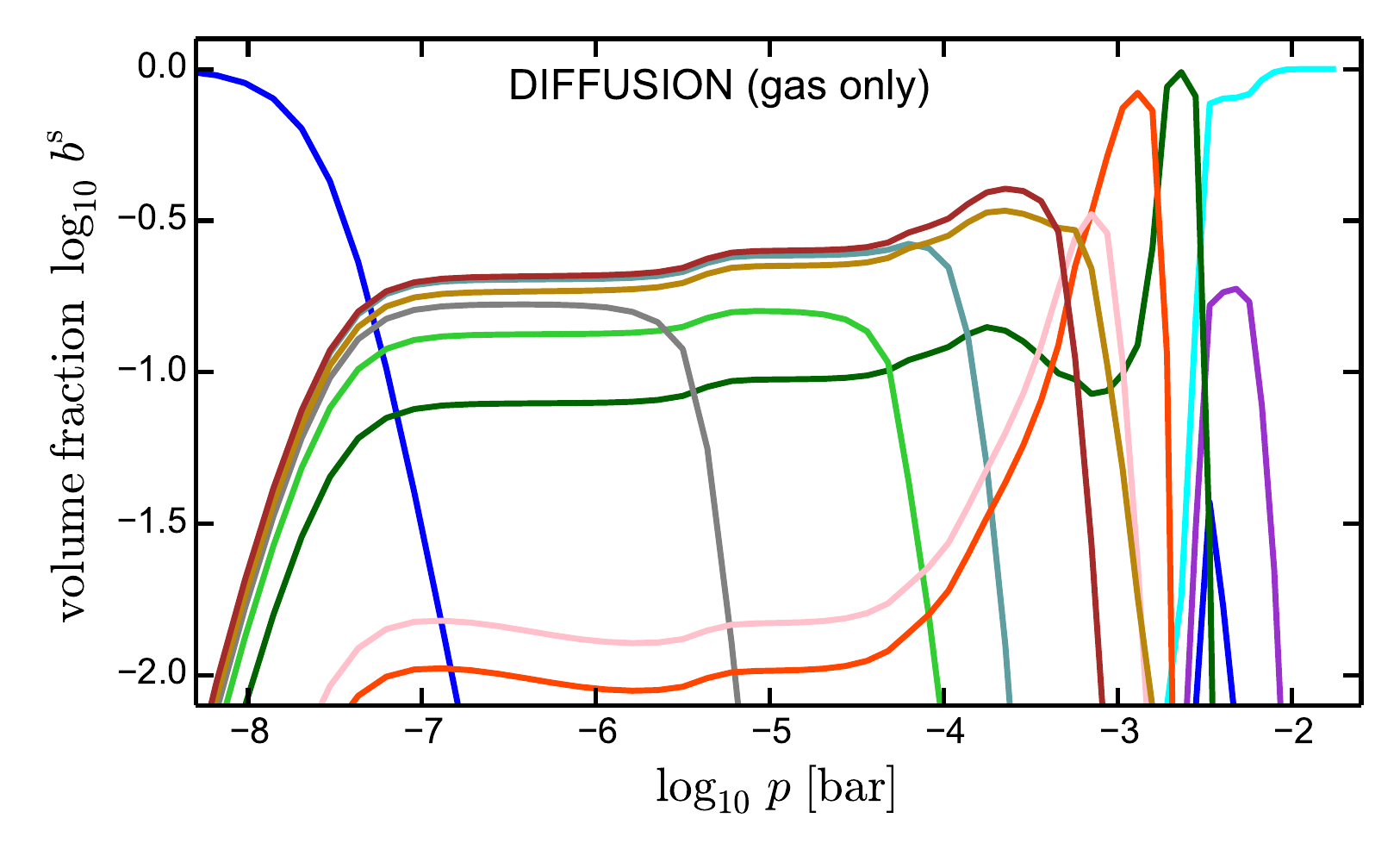}\\[-0.5mm]
  \caption{The material volume composition of the cloud particles
    $b^s\!=\!L_3^{\rm s}/L_3\!=\!V_s/V_{\rm tot}$ for the same three
    models as discussed in Figs.~\ref{fig:oldnew} and
    \ref{fig:oldnew_eps}.}
  \label{fig:VolComp}
  \vspace*{-2mm}
\end{figure} 

\medskip\noindent {\sll Cloud particle composition:\ }
Figure~\ref{fig:VolComp} shows the corresponding solid material
composition (by volume) of the cloud particles. In all three models we
assume that the cloud particles have a well-mixed material composition
which is independent of size, but that composition changes as the particles fall
through the atmosphere, hence material composition depends on
height. All three models show the same basic vertical structure.\pagebreak
\begin{enumerate}
\setlength{\itemsep}{2pt}
\setlength{\parskip}{0pt}
\setlength{\topsep}{-6pt}
\setlength{\parsep}{0pt}
\setlength{\partopsep}{0pt}
\item A layer containing only the most stable metal-oxides at the
  cloud base, in this model Al$_2$O$_3$[s], TiO$_2$[s] and
  CaTiO$_3$[s]. The position of the cloud base, which depends on
  $T_{\rm eff}$ and $\log\,g$, is located at around 1800\,K in this
  model.
\item A thin layer of cloud particles around 1500\,K which mainly
  consist of metallic Fe[s].
\item Main silicate cloud layer composed of Mg$_2$SiO$_4$[s],
  MgSiO$_3$[s], MgO[s], SiO[s] and SiO$_2$[s], mixed with metallic
  iron, upward of about 1400\,K in this model.
\item Less stable solid materials such as FeS[s], FeO[s] and
  Fe$_2$O$_3$[s] are incorporated into the silicate cloud particles at
  temperatures lower than about 1100\,K, 1000\,K and 850\,K,
  respectively, in this model.
\item Pure nuclei at the top, here TiO$_2$[s], which fall through the
  atmosphere so quickly that they practically do not grow.
\end{enumerate}
Further inspection shows, however, that the material composition of
the main silicate cloud layer (3) differs substantially between the
{\sc Drift} and the {\sc DiffuDrift} models. In the new
diffusive {\sc DiffuDrift} models, the first magnesium-silicate to form is fosterite
Mg$_2$SiO$_4$[s], which has a stoichiometric ratio $\rm
Mg\!:\!Si\!=\!2\!:\!1$. The formation process of Mg$_2$SiO$_4$[s]
stops once the reservoir of Mg is exhausted, still leaving about half
of the available Si in the gas phase. Since the mixing is diffusive,
very little Mg can be mixed upwards through these Mg$_2$SiO$_4$[s]
clouds.  Thus, the remaining amount of Si above the Mg$_2$SiO$_4$[s]
clouds preferentially forms other silicate materials, in particular
SiO$_2$[s] and SiO[s], but only very little MgSiO$_3$[s].  This is
different in our previous {\sc Drift} model where the depleted elements are
assumed to be instantly replenished at similar rates, in which case
both Mg$_2$SiO$_4$[s] and MgSiO$_3$[s] are found to be about 
equally abundant condensates in the main silicate cloud layer.

Another difference concerns FeS[s] (troilite).  FeS[s] is found to
form in large quantities in the previous {\sc Drift} models, causing
$\epsilon_{\rm S}$ to drop significantly, see upper part of
Fig.~\ref{fig:oldnew_eps}. However, this depends on our assumptions
about how the elements are replenished.  In the new diffusive models,
upward mixing of gaseous Fe is rather inefficient because the Fe atoms
have plenty of opportunity to condense in form of Fe[s] on existing
cloud particles along their way upwards in the atmosphere.  Once the
temperature is low enough to allow FeS[s] to form, there is so little
Fe left in the gas phase that the S abundance is more or less
unaffected by FeS[s] formation, and therefore sulphur remains
available for other condensates to form.

\subsection{Cloud structures as function of $T_{\rm eff}$}

\begin{figure}
  \vspace*{-1mm}
  \hspace*{-1mm}
  \includegraphics[width=91mm,trim=16 16 6 7,clip]{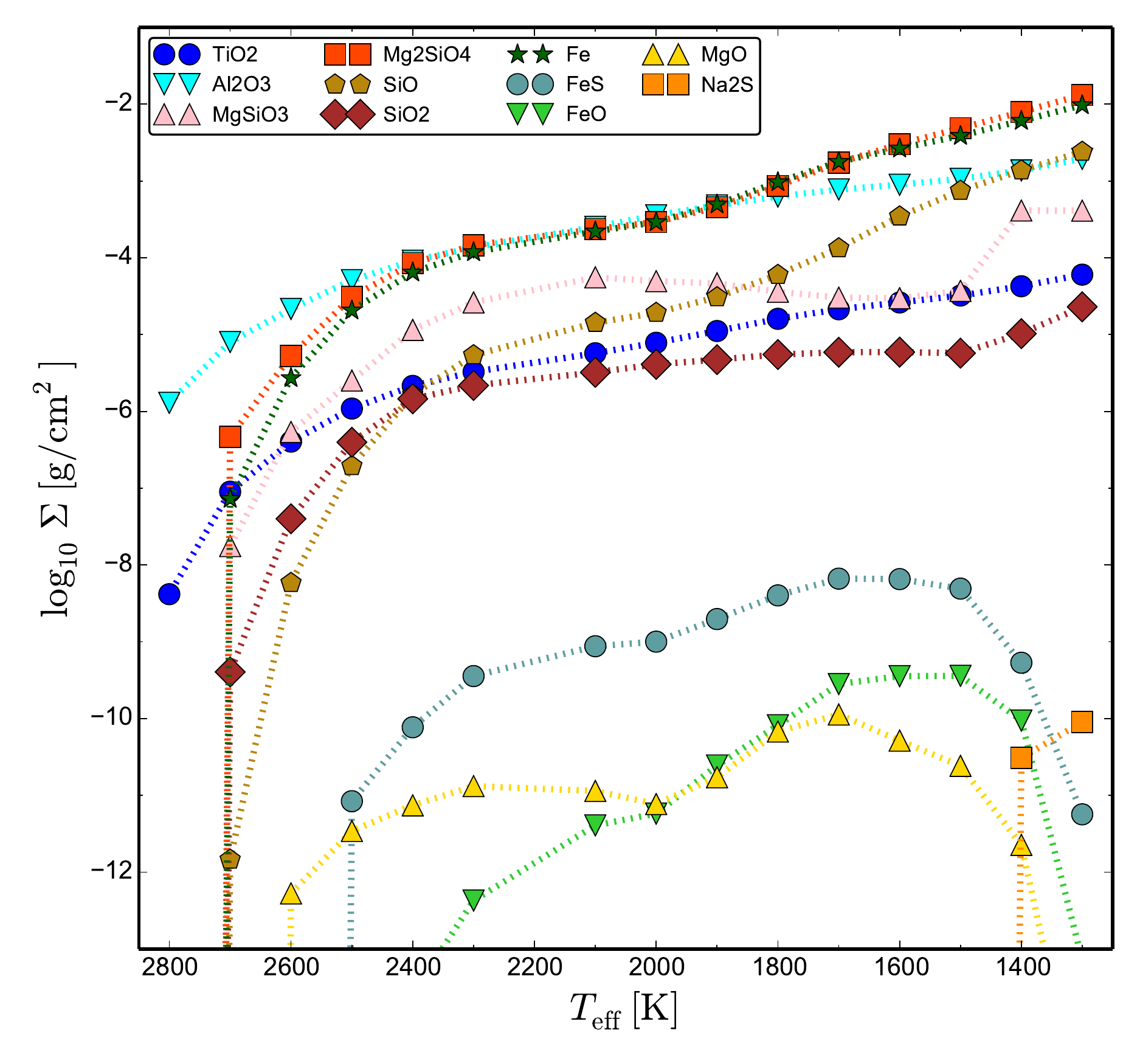}\\[-4mm]
  \caption{Column densities $\rm[g/cm^2]$ of different condensates in
    the atmosphere along a sequence of models with decreasing $T_{\rm
      eff}$ but constant $\log g\!=\!3$ and
    ${\beta^{\prime}}\!=\!1$. A value of $\rm 10^{-3}\,g/cm^2$
      roughly corresponds to an optical depth of one at a wavelength
      of $\lambda\!=\!1\,\mu$m (see Appendix~\ref{sec:opac}).}
  \label{fig:coldens}
\end{figure}

\begin{figure*}
  \begin{tabular}{cc}
  \hspace*{-3mm}
  \includegraphics[width=91mm,height=53mm,trim=10 54 0 12,clip]{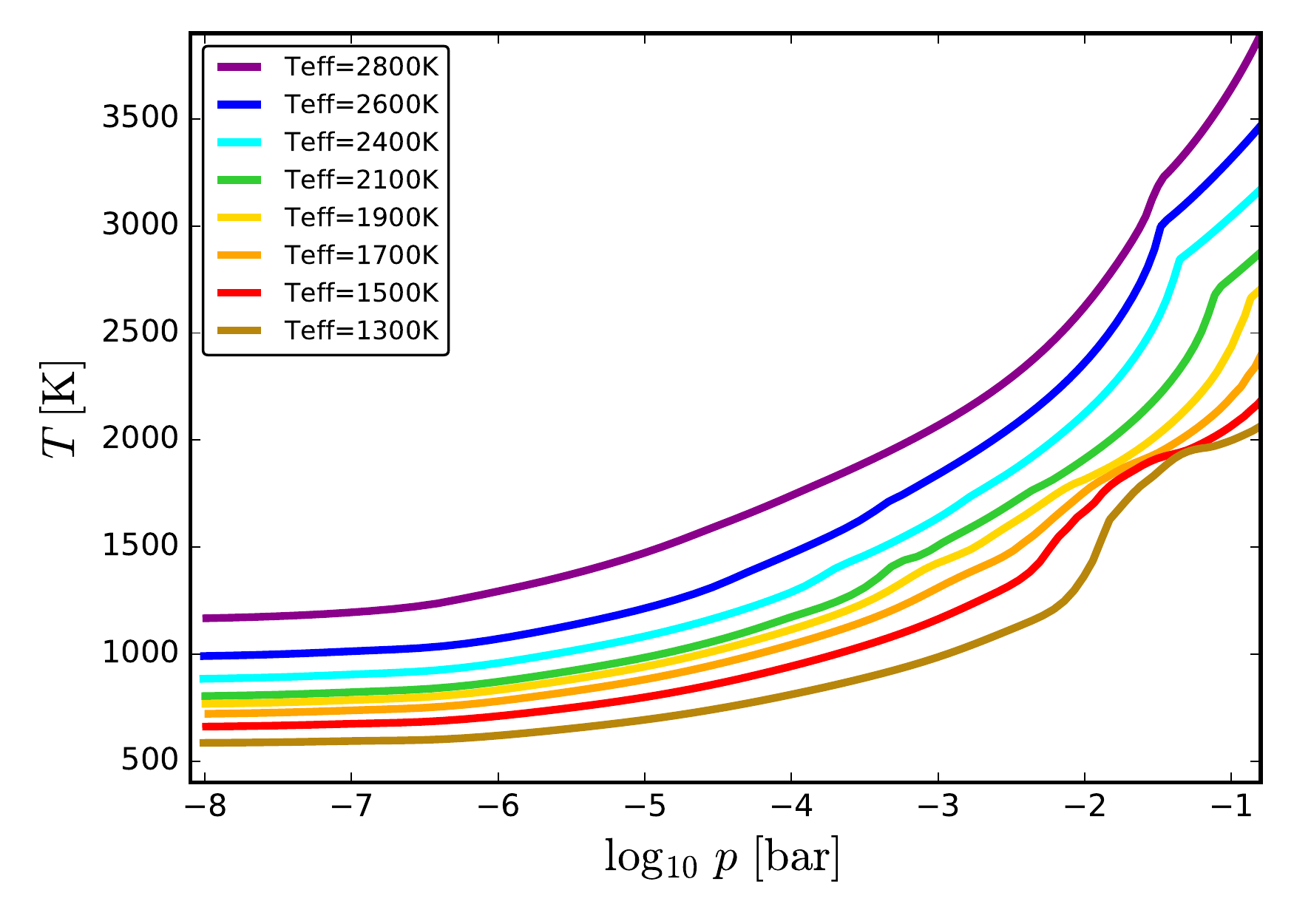} &
  \hspace*{-4mm}   
  \includegraphics[width=91mm,height=53mm,trim=10 54 0 12,clip]{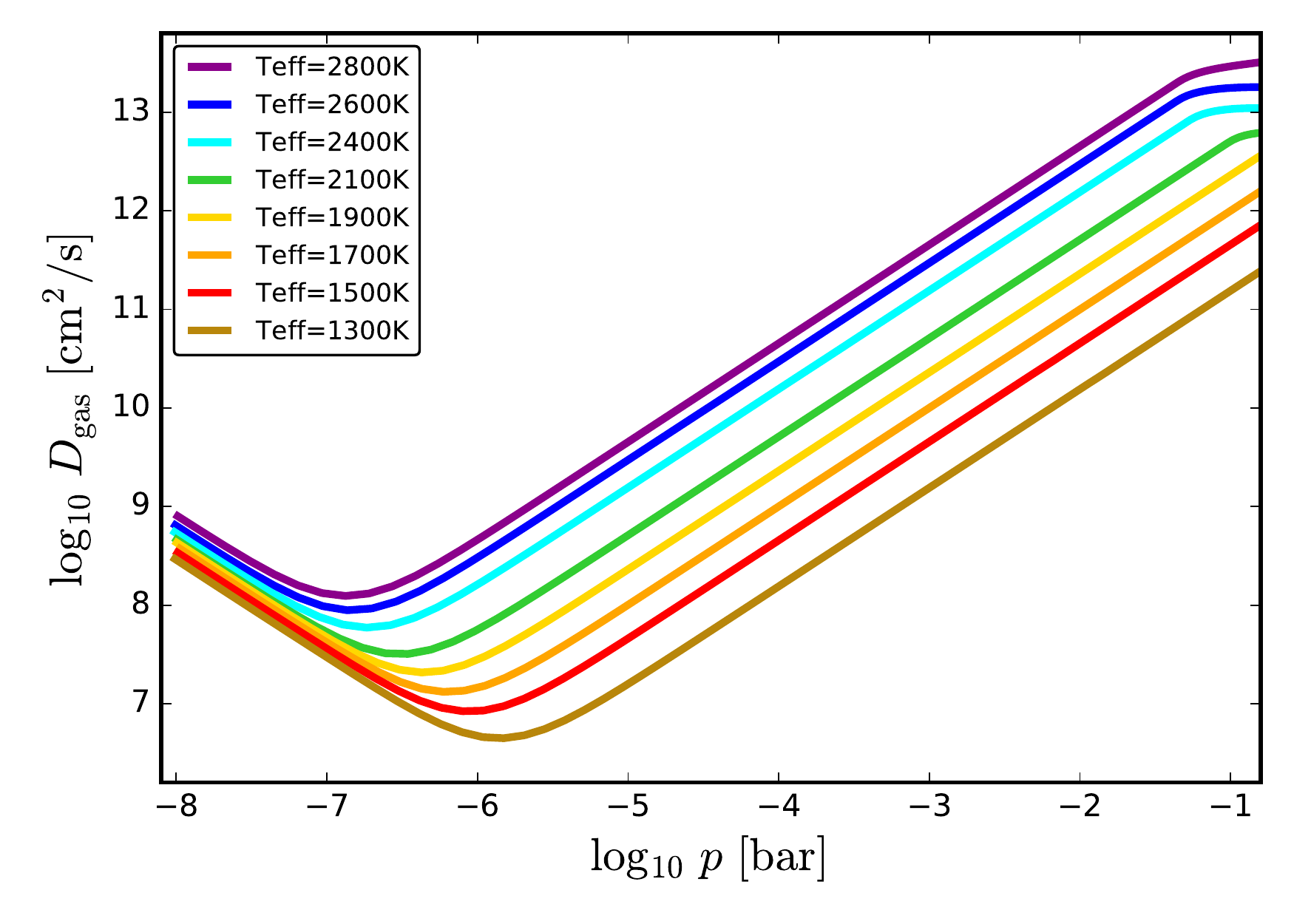}\\[-0.7mm]
  \hspace*{-3mm}
  \includegraphics[width=91mm,height=53mm,trim=-4 54 0 12,clip]{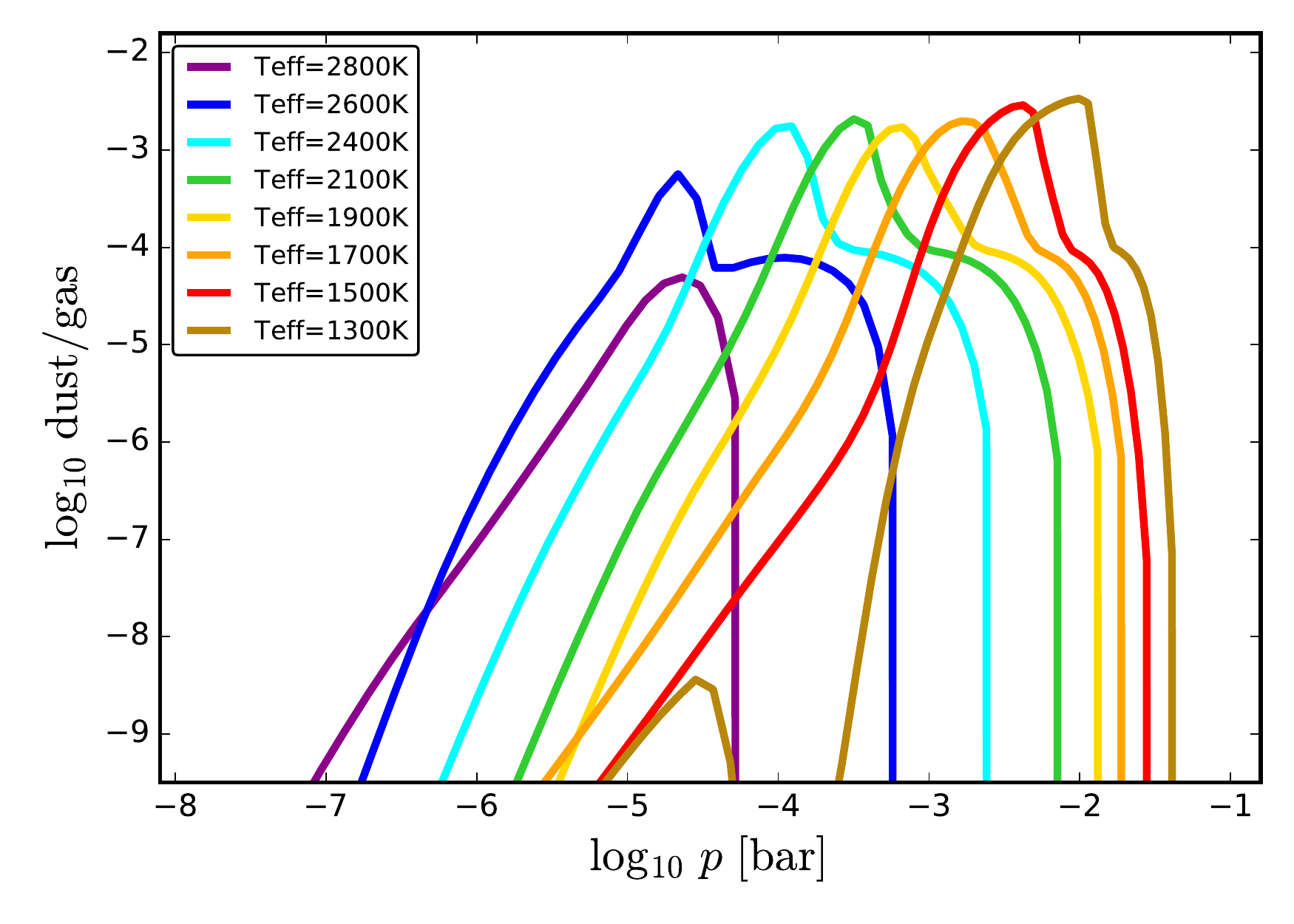} &
  \hspace*{-4mm}   
  \includegraphics[width=91mm,height=53mm,trim=15 54 0 12,clip]{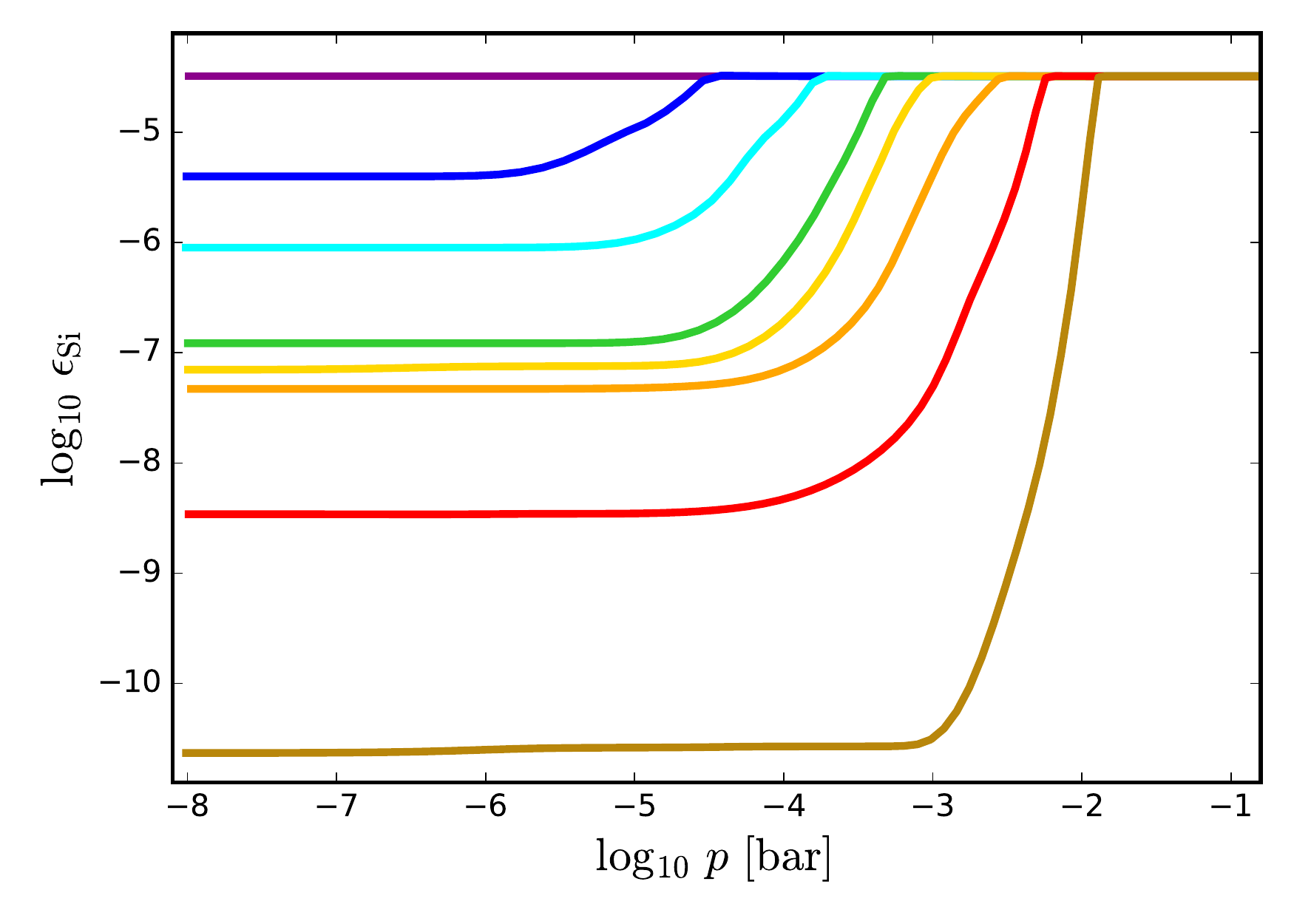}\\[-0.7mm]
  \hspace*{-3mm}
  \includegraphics[width=91mm,height=60mm,trim= 6 15 0 12,clip]{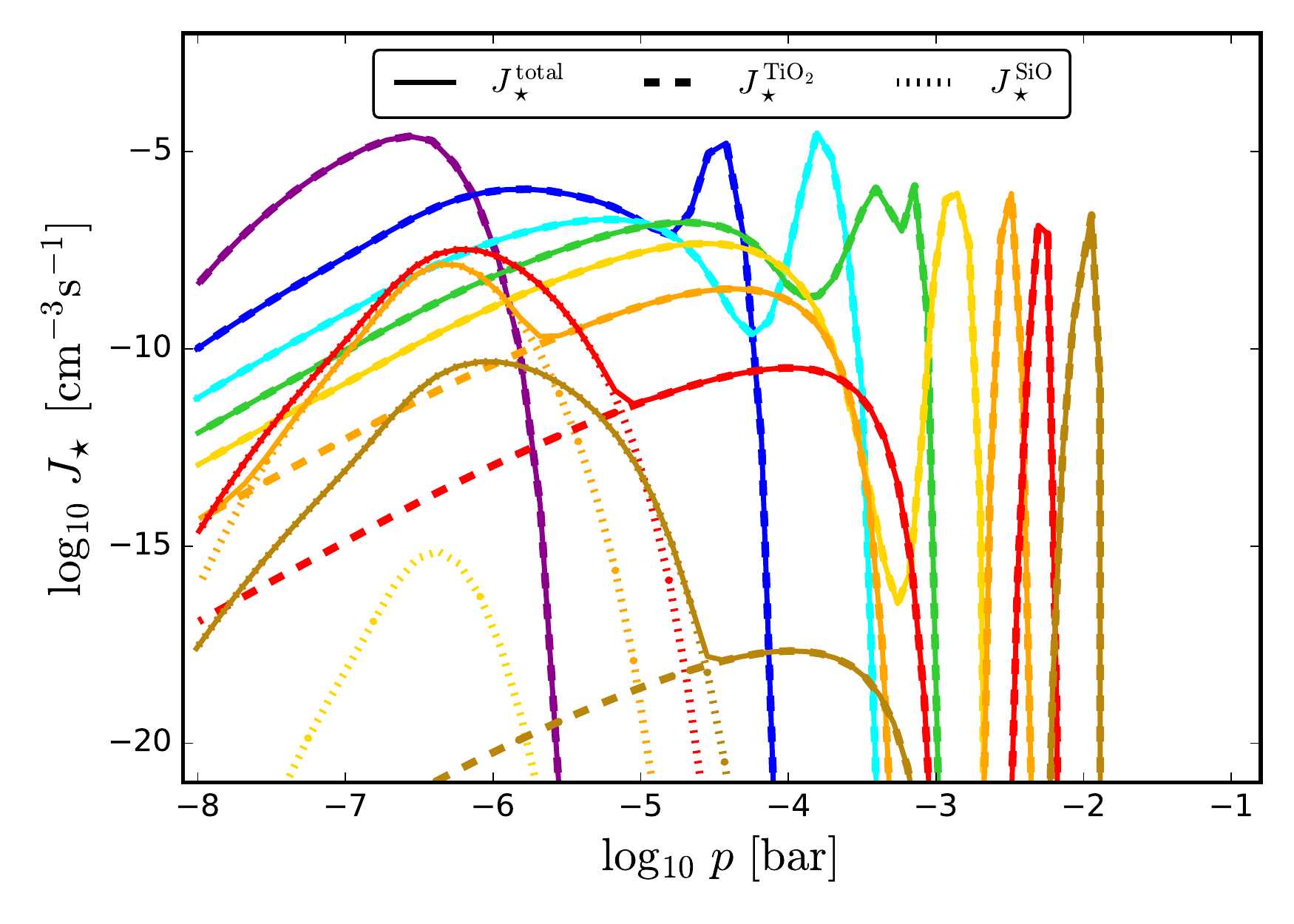} &
  \hspace*{-4mm}   
  \includegraphics[width=91mm,height=60mm,trim=13 15 0 12,clip]{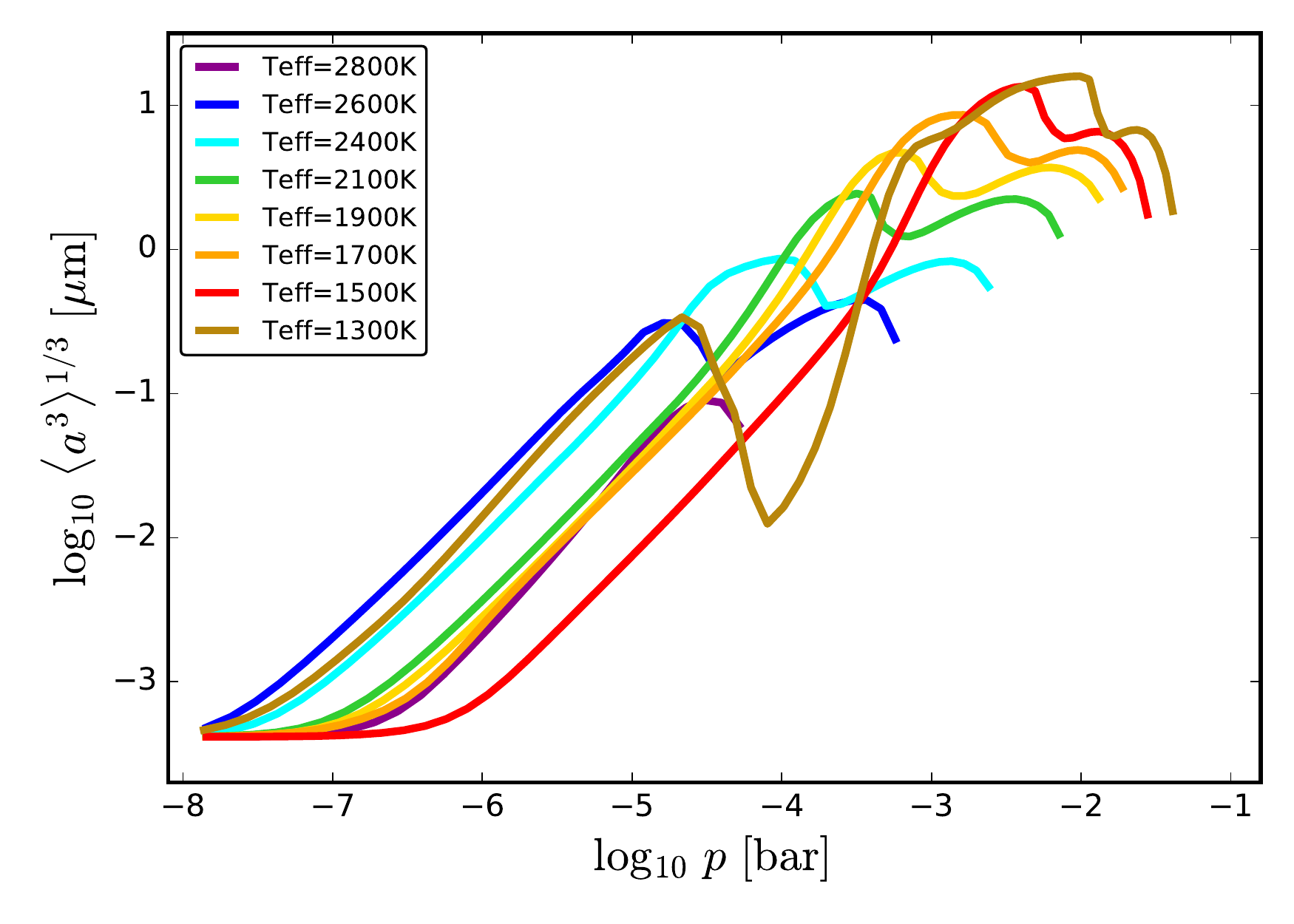}\\[-0.7mm]
  \end{tabular}
  \caption{Sequence of cloud forming models with decreasing $T_{\rm
      eff}$ at constant $\log g\!=\!3$ and mixing powerlaw index
    ${\beta^{\prime}}\!=\!1$. {\bf Top row:} gas temperature $T$
    and diffusion coefficient $D_{\rm gas}$ as function of pressure
    (both assumed).  {\bf Middle low:} resulting dust to gas mass
    ratio and element abundance of silicon in the gas phase
    $\epsilon_{\rm Si}$. {\bf Lower row:} resulting nucleation rates
    $J_\star$ and mean particle sizes $\langle a^3\rangle^{1/3}$.}
  \label{fig:Tseries}
  \vspace*{-2mm}
\end{figure*}

In this section, we study the results of a sequence of the
new {\sc DiffuDrift} cloud formation models with decreasing effective
temperature $T_{\rm eff}$. We are using a slightly different chemical setup here
that will allow us to discuss secondary cloud layers.  We consider
four nucleation species ${\rm(TiO_2)}_N$, ${\rm(SiO)}_N$,
${\rm(KCl)}_N$ and ${\rm(C)}_N$. The nucleation rates of TiO$_2$, KCl
and C are calculated by modified classical nucleation theory
\citep{2017A&A...603A.123H, Gail1984}, with a surface tension value
for KCl from \citep{Lee2018}.  The nucleation rate of SiO
is calculated according to \citep{Gail2013}.  We have 16 elements in
this setup (H, He, Li, C, N, O, Na, Mg, Si, Fe, Al, Ti, S, Cl, K, Ca),
14 condensed species (TiO$_2$[s], Al$_2$O$_3$[s], MgSiO$_3$[s],
Mg$_2$SiO$_4$[s] ,SiO[s], SiO$_2$[s], Fe[s], FeS[s], FeO[s], MgO[s],
KCl[s], NaCl[s], Na$_2$S[s], C[s]), 308 molecules and 50 surface
growth reactions. Molecular equilibrium constants and Gibbs free
energies of the condensates are all taken from \citet{Woitke2018}.
Dust diffusion is included in all models.

Figure~\ref{fig:coldens} shows the total column densities of selected
cloud materials $\Sigma_s\,\rm[g/cm^2]$ in a series of {\sc DiffuDrift}
models with constant $\log g\!=\!3$ and
mixing index ${\beta^{\prime}}\!=\!1$, but decreasing
$T_{\rm eff}$. The column densities of the
condensed species are computed as
\begin{equation}
  \Sigma_s = \int \rho_s\;\rho L_3^s\;dz   \ ,
\end{equation}
where $\rho_s\rm\,[g/cm^3]$ is the material density of the pure
condensate of kind $s$ and $\rho L_3^s\rm\,[cm^3/cm^3]$ is the volume
of condensed kind $s$ per volume of atmosphere. For example, for
$T_{\rm eff}\!=\!1800\,$K we find of order $\rm 10\,mg$ condensates
per square centimetre, mostly made of $\rm Mg_2SiO_4[s]$,
Fe[s] and $\rm Al_2O_3[s]$, followed by SiO[s] and $\rm MgSiO_3[s]$.

On the left side of this plot, the first model that shows condensation
appears at $T_{\rm eff}\!=\!2800\,$K. Here, the temperatures are too
high to have any other condensates than just the most stable
metal-oxides in form of $\rm Al_2O_3[s]$ and $\rm TiO_2[s]$. In the
next few models down to $T_{\rm eff}\!=\!2000\,$K, the main silicate
layer forms, mixed with iron. In this range of effective temperatures,
$\rm Al_2O_3[s]$ still has the largest column density because the
metal oxide layer is situated deeper in the atmosphere where the
densities are higher. Only for $T_{\rm eff}\!<\!2000\,$K, the
silicate-iron layer starts to dominate by mass.  At the very end of
the sequence, for $T_{\rm eff}\!<\!1500\,$K we find the first models
which host a third cloud layer made of di-sodium sulfide $\rm Na_2S[s]$.

Figure~\ref{fig:Tseries} shows a few more details from this $T_{\rm
  eff}$-series of new cloud formation models.  The upper left plot
shows the atmospheric density/temperature structures assumed (taken
from the {\sc Drift-Phoenix} atmosphere grid
\citep{2007IAUS..239..227D, hell2008, Witte2009, 2011A&A...529A..44W}.
The kinks in deep layers ($T\!\sim\!2500-3000\,$K) indicate the
beginning of the convective layer (Schwarzschild
criterion for convective instability).  The upper
right plot shows the assumed diffusion coefficient in the atmosphere,
which decreases with $T_{\rm eff}$, because the convective layer sinks
into deeper layers, hence the spatial distance to the source causing
the mixing motions in the atmosphere increases.

The left middle plot in Fig.~\ref{fig:Tseries} shows the dust-to-gas
mass ratio, which has its maximum in the main silicate-iron layer, and
a shoulder on the right due to the deeper metal-oxide clouds which are
made of the rarer elements with the highest condensation temperatures,
namely aluminium, calcium and {titanium}. As $T_{\rm eff}$
decreases, both layers move inward to deeper layers and become
successively more narrow, until finally, for $T_{\rm
  eff}\!=\!1400\,$K, a new cloud layer occurs which mainly consists of
di-sodium sulfide $\rm Na_2S[s]$. The right middle plot shows how the
silicon abundance in the gas phase is affected. All curves are
monotonic decreasing towards the surface, with higher Si depletions
for lower $T_{\rm eff}$ where the silicate cloud particle formation is
more complete.

The nucleation rates of $\rm (TiO_2)_N$ and $\rm (SiO)_N$ particles
are depicted in the lower left plot. A complicated, double-peaked
pattern shows, which has a minimum around the main peak of the
dust-to-gas ratio (at the peak position of the main silicate-iron
layer).  ${\rm(TiO_2)}_N$ is usually the most significant nucleation
species, but cooler models show additional contributions by
${\rm(SiO)}_N$.  The resulting mean particle sizes are plotted on the
lower right, with a tendency to produce larger particles deep in the
atmosphere for lower $T_{\rm eff}$. An in-between minimum in particle
size occurs where the main silicate material evaporates. Only the
coolest model has a second minimum where $\rm Na_2S[s]$ evaporates.
Interestingly, the hottest and the coolest model in
Fig.\,\ref{fig:Tseries} show about equally large cloud particles at
high altitudes, whereas all other models show smaller particles.

\section{Summary and Discussion}

This paper has introduced a new cloud formation model applicable to
the atmospheres of brown dwarfs and gas giant (exo-)planets.  We have
combined our previous cloud particle moment method \citep{Woitke2004,
  Helling2006, Helling2008} with a diffusive mixing approach,
according to which, in the final relaxed time-independent state of the
atmosphere, fresh condensable elements are diffusively transported
upwards to replenish the upper atmosphere via a combination of
turbulent (eddy) mixing and gas-kinetic diffusion. Our formulation of
the problem arrives at a system of about 30 second order partial
differential equations of reaction-diffusion type, where the formation
and growth of the cloud particles follows from a kinetic treatment in
phase-non-equilibrium.

\begin{figure}
  \vspace*{1.5mm}\hspace*{-1mm}
  \includegraphics[width=91mm,trim=5 2 0 3,clip]{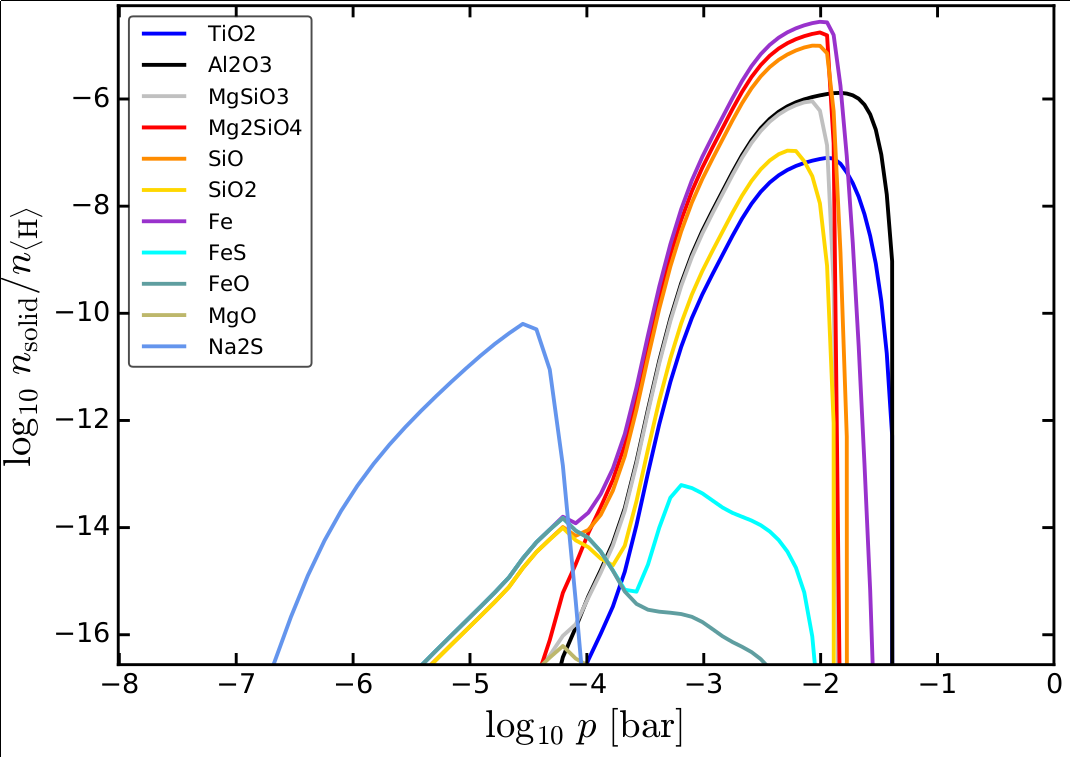}\\[-5mm]
  \caption{Concentration of condensed species in a model with
    $\Teff\!=\!1300\,$K, $\log\,g\!=\!3$ and
    ${\beta^{\prime}}=\!=\!1$, showing a secondary cloud layer
    almost entirely made of di-sodium sulfide Na$_2$S[s].\ \ $n^s_{\rm
        cond}\!=\!\rho L_3^s/V_{\rm mat}^s\rm\,[cm^{-3}]$ is the number
      density of solid units of condensed species $s$, $V_{\rm
        mat}^s\!=\!m_s/\rho_s$ is the volume occupied by one unit of solid
    $s$ in the pure material, and $m_s$ is the mass of one such units,
    for example 100.4\,{\rm amu} for $s\!=\!\rm MgSiO_3$.\ \ $n^s_{\rm
        cond}/\nH$ is directly comparable to element abundances.}
  \label{fig:Na2S}
  \vspace*{-1mm}
\end{figure}

\smallskip\noindent {\sll Model setup:\ } The new cloud formation
model is applied to a given one-dimensional $(p,T)$ atmospheric
structure in this paper.  The model is calculated time-dependently,
using an operator splitting technique.  All models are found to relax
toward a time-independent, stationary solution, where the condensable
elements are constantly mixed up diffusively, cloud particles nucleate
from the gas phase high in the atmosphere, grow by the simultaneous
condensation of different solid materials on their surface, and then
decent through the atmosphere due to gravitational settling, before
the particles stepwise purify and eventually sublimate completely at
the cloud base.

\smallskip\noindent {\sll Timescales:\ } The real-time simulation time
required to reach that stationary solution varies between a few months
to several tens of years, depending on $\log(g)$ and $\Teff$.  The
relaxation is quicker when models are started from an atmosphere that
is devoid of any condensable elements at $t\!=\!0$. These relatively
long simulation times make these models computationally expensive (of
order 500 CPU-hours per model), because the intrinsic nucleation and
growth reactions are very fast, {which means that the models need to
  be advanced on short computational time steps of the order of
  seconds to guarantee numerical stability}.  The long physical
timescales involved in the simulations are (i) the overall settling
time for small particles inserted high in the atmosphere, and (ii) the
overall mixing time for gas parcels to diffusively reach the highest
point in the model from the cloud base.  This implies that 3D
simulations of cloud formation (GCM models, for example
\citealt{2010A&A...513A..19F, 2016A&A...594A..48L,
  Lines2018a,2018ApJ...860...18P, 2018ApJ...854..172C}) must be
advanced for similar real-time simulation times before a relaxed
physical structure can be expected. However, how long these physical
timescales actually are will depend on the exact formulation of mixing
and setting in the GCM models.

\smallskip\noindent {\sll Cloud density and particle sizes:\ } In
comparison to our previous {\sc Drift} models, the {\sc DiffuDrift}
models show fewer but larger cloud particles, which are more
concentrated towards the cloud base. However, the physical properties
of the cloud particles in the main silicate-iron layer towards the
bottom of the clouds (dust to gas mass ratio, particle sizes, {optical
  depth}, chemical composition, etc.) are found to be similar to the
results of the previous models. The dust-to-gas ratio in the main
silicate-iron layer reaches a peak value of about 0.002 to 0.003,
quite independent of $\Teff$, for not too hot models
($\Teff\!>\!2500\,$K). {This is close to the maximum value of 0.0045
  as expected from complete condensation of a gas with solar
  abundances \citep{Woitke2018}}. The physical reason for the stronger
concentration of the cloud particles around the cloud base in the {\sc
  DiffuDrift} models is that the diffusive element replenishment is
less effective for the upper atmosphere, because the molecules
carrying the elements diffusively upwards {have a high probability
  to collide with existing cloud particles on their way up the
  atmosphere}. This effect was not accounted for in the previous
models.

\smallskip\noindent {\sll Element abundances:\ } The concentration of
condensable elements in the gas phase shows a steep decline in the
{\sc DiffuDrift} models above the cloud base, followed by a {\sl
  monotonous decrease} towards a plateau which then continues on that
level toward the upper boundary of the model. This behaviour is
expected in the time-independent relaxed state, because the downward
flux of condensable elements due the falling cloud particles must be
compensated for by an upward diffusive flux of elements in the gas
phase, which requires a negative concentration gradient.  We find the
abundances of the condensable elements high above the cloud layers to
be strongly dependent on effective temperature, in agreement with the
results of 2D radiation hydro-models by
\citet{2010A&A...513A..19F}. For example, the silicon abundance is
reduced by about 2.5 orders of magnitude for $\Teff\!=\!2000\,$K, but
6 orders of magnitude for $\Teff\!=\!1300\,$K in our models.

\smallskip\noindent {\sll Cloud material composition:\ } The chemical
composition of the cloud particles is characterised by (i) a deep
layer with the most stable metal-oxides at the cloud base
(Al$_2$O$_3$[s], TiO$_2$[s], CaTiO$_3$[s] in the current setup), (ii)
the main silicate-iron layer (mainly Mg$_2$SiO$_4$[s] and Fe[s])
which, with increasing height, is then mixed with other silicates like
SiO[s] and SiO$_2$[s] and MgSiO$_3$[s]. Some less stable condensates
are also found in smaller quantities, in particular FeS[s] and FeO[s].
The condensation in both these cloud layers leads to a
removal of certain elements from the gas phase, and the stoichiometry
of the condensates decides upon which elements remain for further
condensation higher in the atmosphere.  In particular, the formation
of Mg$_2$SiO$_4$[s], with stoichiometry $\rm Mg\!:\!Si\!=\!2\!:\!1$
causes the Mg abundance to drop quickly, whereas roughly half of the
originally available Si remains in the gas phase, which then favours
the formation of SiO[s] and SiO$_2$[s] above the Mg$_2$SiO$_4$[s]
layer, rather than the formation of MgSiO$_3$[s], which is a rather
unimportant condensate in our new {\sc DiffuDrift} models.
Having so little MgSiO$_3$[s] in the main silicate-iron layer is a
result that differs from the results obtained with our previous {\sc
  Drift} models, and from phase-equilibrium models starting from
complete solar abundances \citep{Woitke2018}.

\smallskip\noindent {\sll Na$_2$S[s] clouds:\ } Our coolest {\sc
  DiffuDrift} models show the occurrence of a secondary cloud layer
almost entirely made of di-sodium sulfide Na$_2$S[s], see
Fig.\,\ref{fig:Na2S}. The presence of Na$_2$S-clouds in brown dwarf
atmospheres has been proposed by \cite{2012ApJ...756..172M} to fit the
optical appearance of two red T-dwarfs.  The formation of
Na$_2$S-clouds requires the presence of sulphur and sodium in the gas
phase at low temperatures.  In phase equilibrium models starting from
solar abundances, such a combination is prevented by the formation of
FeS[s] (troilite), which consumes the sulphur. However, in our new
diffusive kinetic cloud formation models, iron is depleted by the
formation of metallic iron Fe[s] at high temperatures, so FeS[s]
cannot form in large quantities. Consequently, sulphur remains
available to eventually form Na$_2$S[s] at lower temperatures. The
condensation of Na$_2$S[s] then reduces the possibility to form
NaCl[s] at even lower temperatures, and so on.  Therefore, the new
diffusive {\sc DiffuDrift} models reveal new details about the
condensation sequence in cloudy atmospheres, and we need more
experiments with our selection of condensates during model
initialisation to arrive at more distinct conclusions.

\section{Conclusions}

The physical description of the replenishment mechanism for
condensable elements in planetary atmospheres seems crucial for
realistic cloud formation models. This paper has used a
quasi-diffusive approach in 1D to simulate the turbulent eddy-mixing
processes in cloudy atmospheres, using the new {\sc DiffuDrift}
models. This approach can be considered as the limiting case of
small-scale mixing. On the other extreme, large-scale hydrodynamic
motions (convection, Hadley-cells, etc.)  may be able to dredge up
those elements maybe in a more immediate straightforward way, which
was the idea in our previous {\sc Drift} models.  In reality, there is
not only vertical, but also horizontal mixing, which is likely to be
very efficient for example in super-rotating horizontal jets as known
from Jupiter \citep{Schneider2009}, assuming that there are
considerable horizontal abundance gradients present in the
atmosphere. More 3D numerical experiments are required to quantify the
efficiency of mixing to inform our cloud formation models.

\begin{acknowledgements}
      ChH thanks Will Best and Jonathan Gagn{\'e} for the discussion
      on the number of known brown dwarfs. We thank Robin Baeyens and
      Ludmila Carone for insightful discussions on mixing regimes in
      giant gas planets. The computer simulations were carried out on
      the UK MHD Consortium parallel computer at the University of St
      Andrews, funded jointly by STFC and SRIF.
\end{acknowledgements}

\bibliographystyle{aa}
\bibliography{references}


\begin{appendix} 

\section{The diffusion solver}
\label{app:diff}

We use a self-developed 1D explicit/implicit diffusion solver in this
paper which has second order accuracy in both the formulation of the
differential equations and the boundary conditions\footnote{The code
  is available at \url{https://github.com/pw31/Diffusion}.}. In
case of a plane-parallel static atmosphere, the diffusion equation for
element $k$ is given by
\begin{equation}
  \frac{d(\nH\,\ek)}{dt} ~=~ 
     \frac{d}{dz}\left(\nH D_{\rm gas}\frac{d\ek}{dz}\right)
  \label{basic}
\end{equation}
where the diffusive element flux is
\begin{equation}
  j^{\,\rm diff}_k ~=~ -\,\nH D_{\rm gas}\frac{d\ek}{dz}
\end{equation}

\subsection{Vertical grid and discretisation of derivatives}
We introduce an ascending vertical grid $z_i\ (i=1, ...\,,I)$. The first and
second derivatives of any quantity $f(z_i)=f_i$ at grid point $z_i$ 
are approximated as
\begin{eqnarray}
  \frac{df_i}{dz}     
  &=& d^{\,l,1}_if_{i-1} + d^{\,m,1}_if_i + d^{\,r,1}_if_{i+1} \\ 
  \frac{d^2f_i}{dz^2} 
  &=& d^{\,l,2}_if_{i-1} + d^{\,m,2}_if_i + d^{\,r,2}_if_{i+1} \ ,
\end{eqnarray}
i.e.\ as linear combinations of the function values on the
neighbouring grid points, where e.g.\ $d^{\,l,1}_i$ is the coefficient
for the first derivative on the point left of the grid point $i$,
$d^{\,m,1}_i$ the same on the mid point and $d^{\,r,1}_i$ the same on
the point right of grid point $i$. Similar, for the second derivative,
the coefficients are $d^{\,l,2}_i$, $d^{\,m,2}_i$ and
$d^{\,r,2}_i$. Using a second order polynomial approximation for
function $f(z)$ the coefficients are given by
\begin{eqnarray}
  d^{\,l,1}_i &=& -\,\frac{h^r_i}{(h^r_i+h^l_i)\,h^l_i}   \\
  d^{\,m,1}_i &=& +\,\frac{h^r_i-h^l_i}{h^l_i h^r_i}     \\
  d^{\,r,1}_i &=& +\,\frac{h^l_i}{(h^r(i)+h^l_i)\,h^r_i}  \\
  d^{\,l,2}_i &=& +\,\frac{2}{(h^r_i+h^l_i)\,h^l_i}      \\
  d^{\,m,2}_i &=& -\,\frac{2}{h^r_i h^l_i}             \\
  d^{\,r,2}_i &=& +\,\frac{2}{(h^r_i+h^l_i)\,h^r_i}      
\end{eqnarray}   
where $h^l_i=z_i-z_{i-1}$ and $h^r_i=z_{i+1}-z_i$ are the l.h.s.\
and the r.h.s.\  grid point distances. For the special case of
an equidistant grid, we have $h=h^l_i=h^r_i$ and hence
\begin{eqnarray}
  \frac{df_i}{dz}     &=& \frac{f_{i+1}-f_{i-1}}{2h} \\
  \frac{d^{\,2}\!f_i}{dz^2} &=& \frac{f_{i+1}-2f_i+f_{i-1}}{h^2}
\end{eqnarray}
The above equations are valid for grid points $i=2,\,...\,,I-1$.
For the first derivative at the boundaries we write
\begin{eqnarray}
  \frac{df_1}{dz}     
  &=& d^{\,l,1}_1f_{1} + d^{\,m,1}_1f_{2} + d^{\,r,1}_1f_{3} \\ 
  \frac{df_1}{dz}     
  &=& d^{\,l,1}_If_{I-2} + d^{\,m,1}_If_{I-1} + d^{\,r,1}_If_{I} 
\end{eqnarray}
which is also second order accuracy by using the information on the 
3 leftmost or 3 rightmost grid points, respectively.
The coefficients are given by
\begin{eqnarray}
      d^{\,l,1}_1 &=& -\frac{h_2+h_3}{h_2 h_3} \\
      d^{\,m,1}_1 &=&  \frac{h_3}{h_2(h_3-h_2)} \\
      d^{\,r,1}_1 &=& -\frac{h_2}{h_3(h_3-h_2)} \\
      d^{\,r,1}_I &=&  \frac{h_{I-1}+h_{I-2}}{h_{I-1} h_{I-2}} \\
      d^{\,m,1}_I &=& -\frac{h_{I-2}}{h_{I-1}(h_{I-2}-h_{I-1})} \\
      d^{\,l,1}_I &=&  \frac{h_{I-1}}{h_{I-2}(h_{I-2}-h_{I-1})}
\end{eqnarray}
where $h_2=z_2-z_1$, $h_3=z_3-z_1$, $h_{I-1}=z_I-z_{I-1}$ and 
$h_{I-2}=z_I-z_{I-2}$.

\subsection{Spatial derivatives}

The diffusion term at grid point $z_i\;(i=2\,...\,I-1)$ 
is numerically resolved, with abbreviation $D_{\rm gas}(z_i)=D_i$,  as
\begin{eqnarray}
  \left.
  \frac{d}{dz}\left(\nH D_{\rm gas}\frac{d\ek}{dz}\right)\right|_{z_i} 
  = \left.\frac{d\big(\nH D_{\rm gas}\big)}{dz}\right|_{z_i}
  \cdot \left.\frac{d\ek}{dz}\right|_{z_i}
  + \left.\nH D_{\rm gas}\frac{d^2\ek}{dz^2}\right|_{z_i} 
  \hspace*{-8mm}\nonumber\\ 
  = \left(d^{\,l,1}_i\nHim1 D_{i-1} 
          + d^{\,m,1}_i\nHi   D_i 
          + d^{\,r,1}_i\nHip1 D_{i+1}\right) \nonumber\\
   \!\!\!\cdot \left(d^{\,l,1}_i\epsilon_{k,i-1} 
                   + d^{\,m,1}_i\epsilon_{k,i} 
                   + d^{\,r,1}_i\epsilon_{k,i+1} \right) \nonumber\\
  + \nHi D_i \left(d^{\,l,2}_i\epsilon_{k,i-1} 
                   + d^{\,m,2}_i\epsilon_{k,i} 
                   + d^{\,r,2}_i\epsilon_{k,i+1} \right)
  \label{Acoeff}
\end{eqnarray}
and the diffusive fluxes across the lower and upper boundaries are
\begin{eqnarray}
 \phi_{k,1} &=& -D_1\nHleft \left(d^{\,l,1}_1\epsilon_{k,1} 
                            + d^{\,m,1}_1\epsilon_{k,2} 
                            + d^{\,r,1}_1\epsilon_{k,3}\right)\\
 \phi_{k,I} &=& -D_I\nHright \left(d^{\,l,1}_I\epsilon_{k,I-2} 
                            + d^{\,m,1}_I\epsilon_{k,I-1} 
                            + d^{\,r,1}_I\epsilon_{k,I}\right) \ .
\end{eqnarray}

\subsection{Boundary conditions}

As boundary conditions, we have implemented three options, for example
considering the lower boundary: 
\begin{enumerate}
\item fixed concentration: ~~$\epsilon_{k,1}$ is a given constant
\item fixed flux: ~~$\phi_{k,1}$ is a given constant
\item fixed outflow rate: ~~The flux through a boundary is assumed to be
  proportional to the concentration of species $k$ at the boundary,
  e.g.
  \begin{equation}
    \phi_{k,1} = \beta_k\,\nHleft\,\epsilon_{k,1}\,{\rm v}_k  
    \quad\quad{\rm[cm^{-2}s^{-1}]}
  \end{equation}      
  where the $\beta_k$ is a given probability (fixed value) and 
  ${\rm v}_k$ is the speed at which the particles of kind $k$
  are moving through the boundary (also fixed value).
\end{enumerate} 

\subsection{Explicit integration}
A straightforward way to integrate Eq.~(\ref{basic}),
for a timestep $\Delta t$, is the following explicit scheme
\begin{equation}
  f_i^{\,(n)} ~=~ f_i^{\,(n-1)} + \Delta t\,\frac{df_i^{\,(n-1)}}{dt}  
\end{equation}
where $f^{\,(n)}_i$ is some quantity on grid point $i$ at time $t^{\,n}$
and $f^{\,(n-1)}_i$ is the quantity on grid point $i$ at time $t^{\,n-1}$
with $t^{\,n}\!=\!t^{\,n-1}\!+\!\Delta t$. In consideration of Eq.~(\ref{basic}),
this leads to
\begin{eqnarray}
  \epsilon_{k,i}^{(n)} &=& \epsilon_{k,i}^{(n-1)} + \Delta t 
     \sum\limits_{j=1}^{I} A_{ij}\epsilon_{k,j}^{(n-1)} \ ,
  \label{explicit}
\end{eqnarray}
where $\mathbf{A}$ is a tri-diagonal matrix, the elements $A_{ij}$
of which are given by Eq.~(\ref{Acoeff}). Equation~(\ref{explicit})
applies to the grid points $i=2,\,...\,,I-1$, but not to the boundaries. 
On the boundary points, the following equations are applied
depending on the choice of boundary conditions, here for example the
lower boundary
\begin{enumerate}
\item fixed concentration:
      ~~$\displaystyle\epsilon_{k,1}^{(n)} = \ek^0$\\[-1ex]
\item fixed flux:
      ~~$\displaystyle\epsilon_{k,1}^{(n)} = \frac{1}{d^{\,l,1}_1}
       \left(-\frac{\phi_{k,1}}{D_1\nHleft} 
             -d^{\,m,1}_1\epsilon^{(n)}_{k,2} 
             -d^{\,r,1}_1\epsilon^{(n)}_{k,3}\right)$\\
\item fixed outflow rate: 
      \begin{eqnarray*} 
      \beta_k\,\nHleft\,\epsilon_{k,1}\,{\rm v}_k  
      &\!\!=\!\!& -D_1\nHleft\left(
              d^{\,l,1}_1\epsilon_{k,1} 
             +d^{\,m,1}_1\epsilon_{k,2} 
             +d^{\,r,1}_1\epsilon_{k,3}\right)\\
    \Rightarrow\quad \epsilon_{k,1}^{(n)} &\!\!=\!\!&
       \frac{-\,d^{\,m,1}_1\epsilon^{(n)}_{k,2} 
             -d^{\,r,1}_1\epsilon^{(n)}_{k,3}}
            {d^{\,l,1}_1+\frac{\beta_k\,{\rm v}_k}{D_1}}
  \ .
      \end{eqnarray*}
\end{enumerate} 
These assignments are applied at time $t^n$, i.e.\ after an explicit
diffusion timestep has been completed on grid points $i=2\,...\,I-1$.
To guarantee numerical stability, the explicit timestep must be 
limited by $\alpha\leq0.5$ according to
\begin{equation}
  \Delta t ~=~ \alpha\cdot\min_{i\,=\,2,\,...\,,I}
               \frac{(z_i-z_{i-1})^2}{\frac{1}{2}(D_i+D_{i-1})} \ .
  \label{tstep}
\end{equation}

\subsection{Implicit integration}
To avoid the timestep limitations in the explicit solver, and
to guarantee numerical stability for much larger timesteps, an
implicit integration scheme can optionally be applied
\begin{eqnarray}
  f_i^{\,(n)} &=& f_i^{\,(n-1)} + \Delta t \frac{df_i^{\,(n)}}{dt}  
\end{eqnarray}
which is a system of linear equations for the unknowns $f_i^{\,(n)}$.
In consideration of Eq.~(\ref{basic}), we have
\begin{eqnarray}
  \epsilon_{k,i}^{(n)} &=& \epsilon_{k,i}^{(n-1)} + \Delta t 
     \sum\limits_{j=1}^{I} A_{ij}\epsilon_{k,j}^{(n)}
\end{eqnarray}
We re-write this equation more generally, including the
boundary conditions, by means of another unit-free matrix as 
\begin{eqnarray}
  \mathbf{B}\,\vec{\epsilon}_k^{\,(n)} &=& \vec{R}_{k} \ ,
  \label{implicit}
\end{eqnarray}
where we have
\begin{equation}
  B_{ij} = \left(\Eins-\Delta t\,\mathbf{A}\right)_{ij} 
  \quad\mbox{and}\quad R_{k,i} = \epsilon_{k,i}^{(n-1)} 
  \quad\mbox{for $i=2\,...\,I\!-\!1$}
\end{equation}
and, depending on boundary conditions, for example at the lower boundary
\begin{enumerate}
\item fixed concentration: 
      ~$B_{11}\!=\!1$ 
      and $R_{k,1}\!=\!\ek^0$
\item fixed flux:
      ~$B_{11}\!=\!1$, 
      $B_{12}\!=\!{d^{\,m,1}_1}/{d^{\,l,1}_1}$,
      $B_{13}\!=\!{d^{\,m,1}_1}/{d^{\,l,1}_1}$, and\linebreak
      $R_{k,1}\!=\!-\phi_{k,1}/\big(\nHleft D_1 d^{\,l,1}_1\big)$
\item fixed outflow rate:
      ~$B_{11}\!=\!1+\beta_k\,{\rm v}_k/\big(D_1 d^{\,l,1}_1\big)$, 
      $B_{12}\!=\!{d^{\,m,1}_1}/{d^{\,l,1}_1}$,
      $B_{13}\!=\!{d^{\,m,1}_1}/{d^{\,l,1}_1}$, and $R_{k,1}\!=\!0$  \ .
\end{enumerate}
We can now perform an implicit timestep according to Eq.~(\ref{implicit}) 
as
\begin{equation}
  \vec{\epsilon}_k^{\,(n)} ~=~ \mathbf{B}^{-1}\,\vec{R}_{k}
  \label{implicit2}
\end{equation}
where $\mathbf{B}^{-1}$ is the inverse of the matrix $\mathbf{B}$.  As
long as the spatial grid points $z_i$, the densities $\nHi$ and
diffusion constants $D_i$, the constants involved in the boundary
conditions (e.g. $\phi_{k,1}$ or $\beta_k$), and the timestep $\Delta t$
do not change, we need to perform the matrix inversion only
once. Successive time steps are then performed by simply incrementing
$n$, re-computing the vector $\vec{R}_{k}$, and applying again 
Eq.~(\ref{implicit2}). $\mathbf{B}^{-1}$ is also usually the same
for all elements $k$ to be diffused.

This favourable property of $\mathbf{B}$ makes the computation
of implicit timesteps actually very fast. We note, however, that
$\mathbf{B}^{-1}$, in general, is a full $I\times I$ matrix where all
entries are positive $(\mathbf{B}^{-1})_{ij}>0$. This leads to a very
stable numerical behaviour for arbitrary time steps. In contrast, the
matrix $\mathbf{A}$ has positive entries along the main diagonal, but
negative entries along both semi-diagonals, which leads to numerical
instabilities when the time step is too large.

\begin{figure}
\centering
\vspace*{-0.5mm}
\hspace*{-3mm}
\includegraphics[width=94mm]{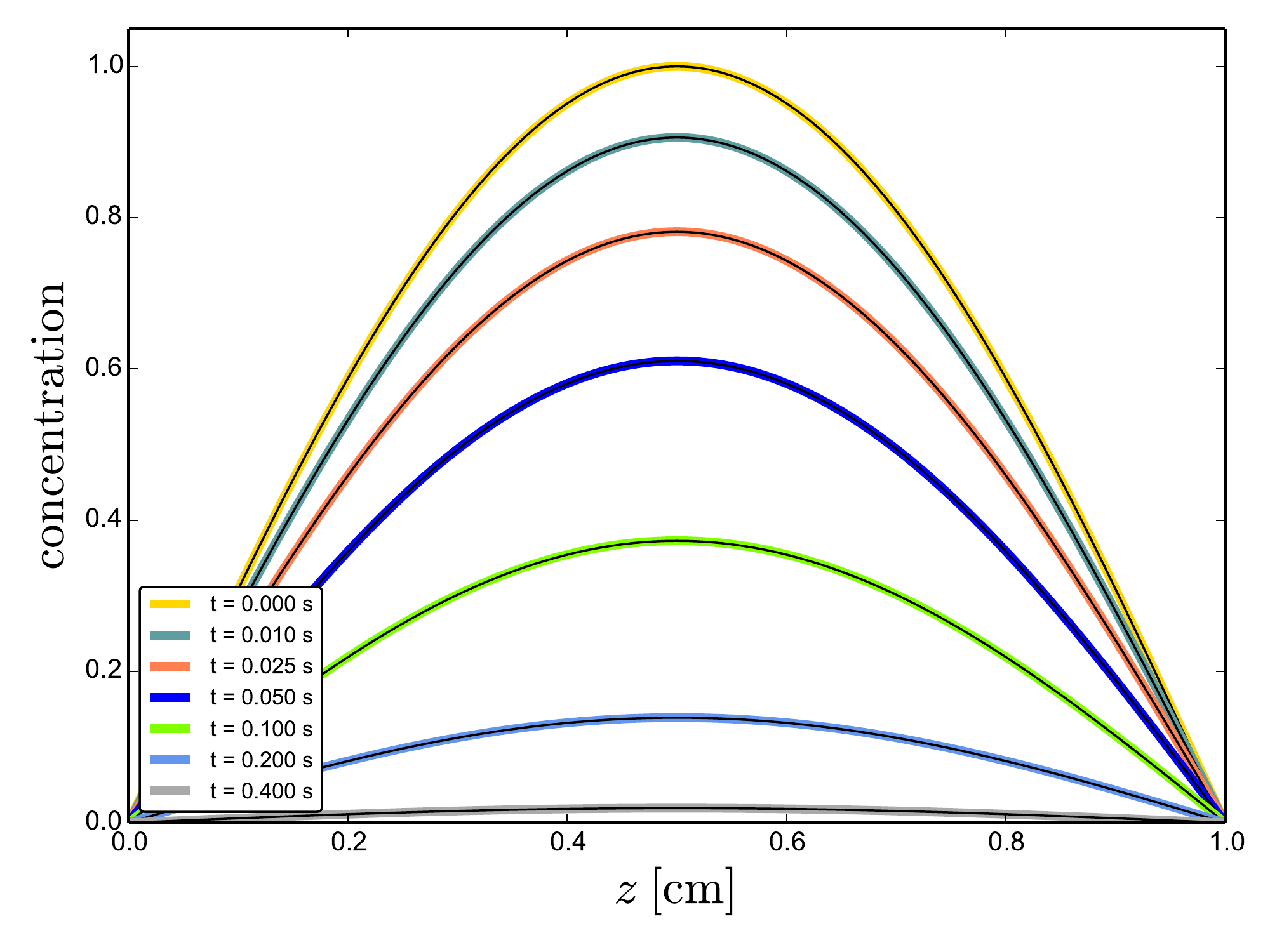}\\[-5mm]
\caption{Test problem with fixed concentrations on the left and right
  boundaries $(\epsilon=0)$. The thin black lines overplot the analytic
  solution $\epsilon(z,t)=\exp(-\omega t) \sin(k z)$ with $k=\pi$ and
  $\omega=D k^2$.}
\label{testfig3}
\end{figure}

\begin{figure}
\centering
\vspace*{-3mm}
\hspace*{-3mm}
\includegraphics[width=94mm]{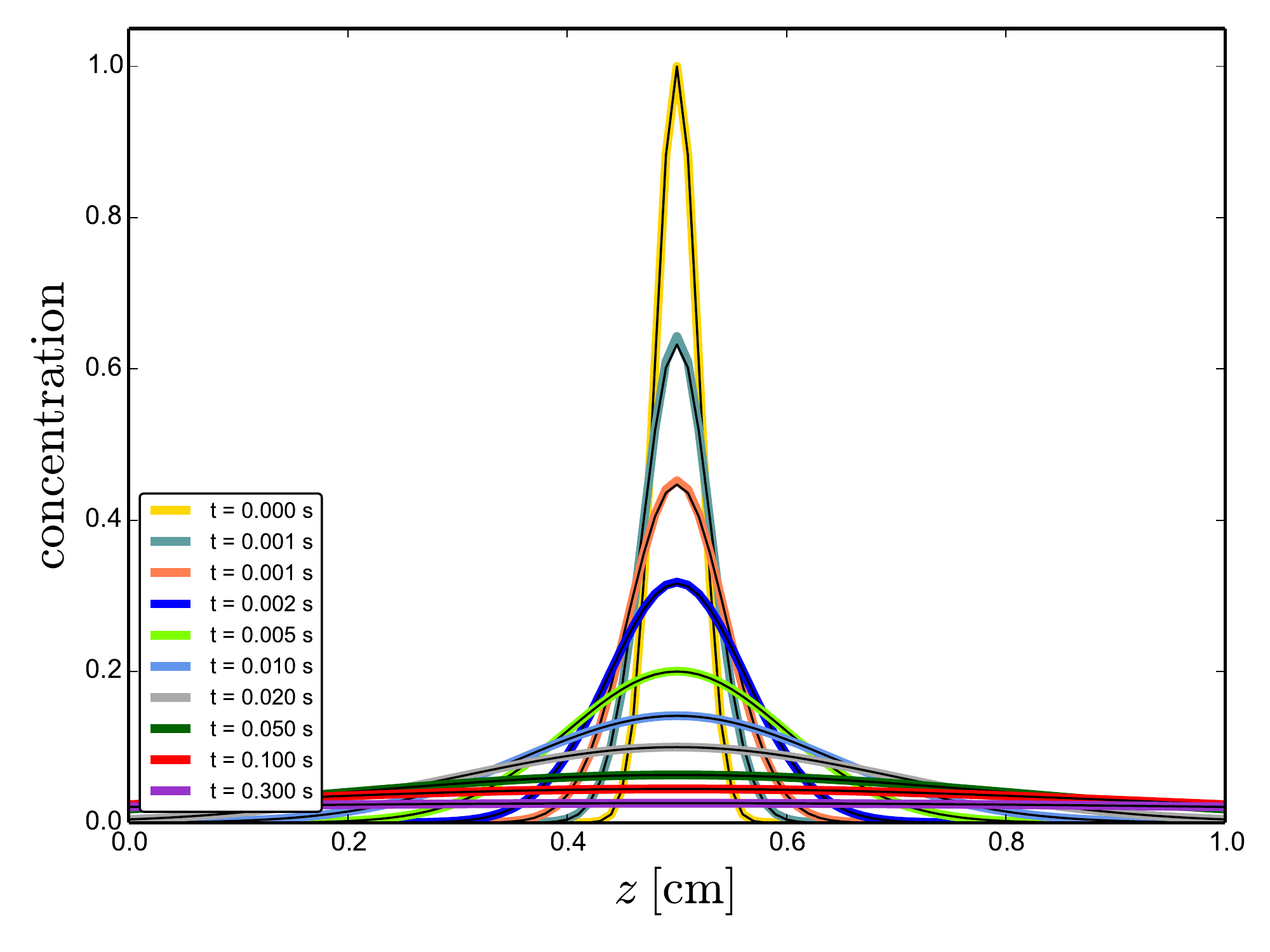}\\[-5mm]
\caption{Diffusive evolution of an initial $\delta$-peak with
  analytic solution overplotted. The analytic solution is
  $\epsilon(z,t)=A(t)\exp\big(-(z\!-\!0.5)^2/w^2(t)\big)$ with
  $A(t)=\sqrt{t_0/t}$ and $w(t)=2\sqrt{D t}$.}
\label{testfig4}
\end{figure}

\begin{figure*}
\centering
\vspace*{-1mm}
\includegraphics[width=164mm]{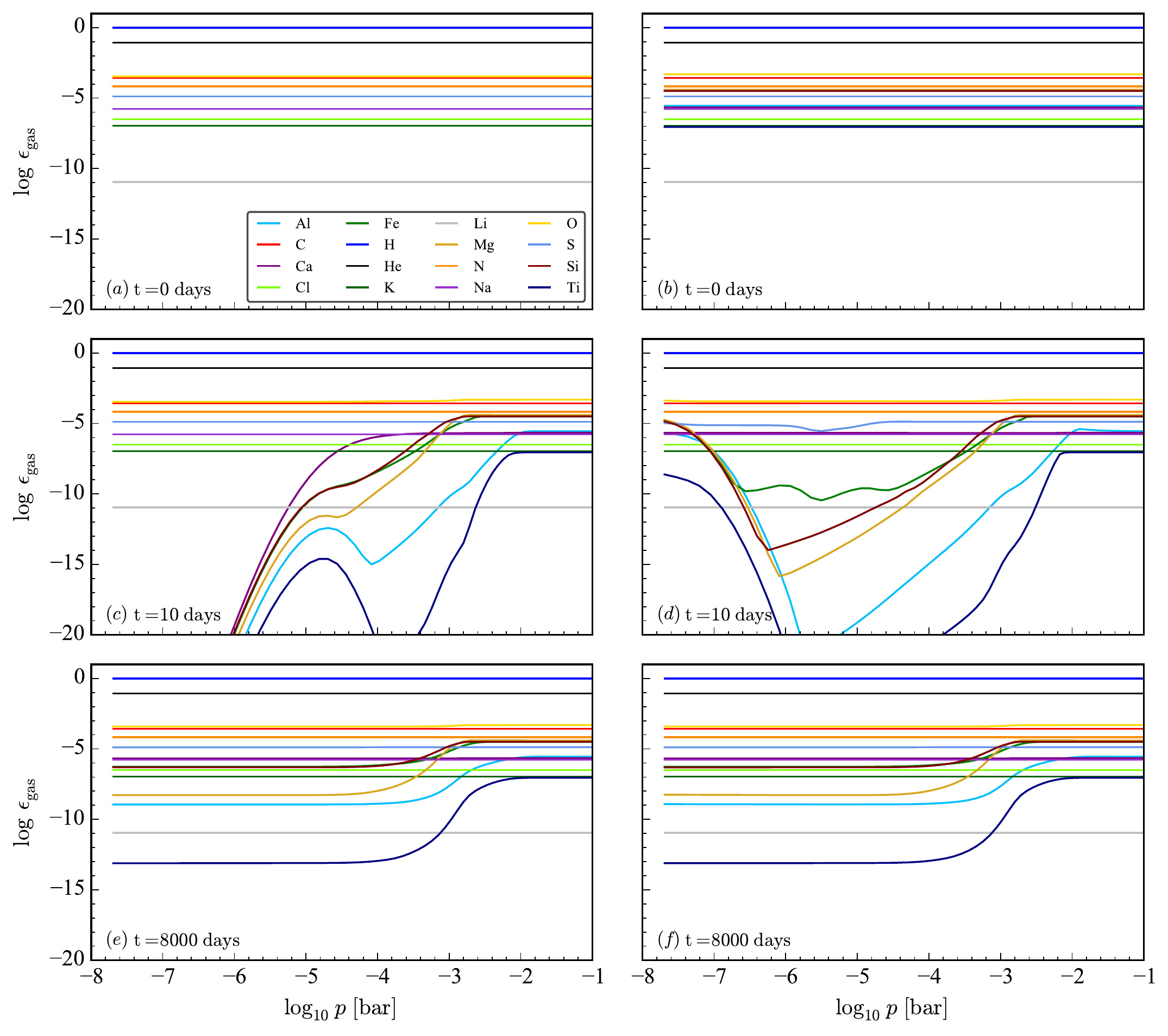}
\vspace*{-3mm}
\caption{{Testing the model convergence for different initial conditions:}
  Gas element abundances $\ek$ as function of pressure $p$ in two
  cloud formation models with different initial conditions 'empty'
  (left) and 'full' (right), see text for further explanations.
  Panels from top to bottom show the respective results after
  $t\!=\!0$, $t\!=\!10$\,days, and $t\,=\,8000$\,days.}
\label{fig:init}
\vspace*{-1mm}
\end{figure*}

\section{The settling solver}
\label{app:sett}

For the 1D vertical settling in the Epstein regime we solve
\begin{equation}
  \frac{d(\rho L_j)}{dt} ~=~ 
  \frac{d}{dz}\int_{\Vl} {\rm v}_{\rm dr}^{\hspace{-0.9ex}^{\circ}}(V)
                        \,f(V)\,V^{j/3}\,dV ~=~ 
  \frac{d}{dz}\left(\xi\,\frac{\rhod}{c_T} L_{j+1}\right) \ ,
\end{equation}
according to Eqs.\,(\ref{eq:vdr1}) and (\ref{eq:xi}).
The settling flux for cloud particle moment $\rho L_j$ is hence
\begin{equation}
  \phi_j = -\,\xi\,\frac{\rhod}{c_T} L_{j+1} 
      = -\,\rho L_j\;{\rm v}_{{\rm dr},j}
  \label{eq:jj}
\end{equation}
where we have introduced mean drift velocities for
the cloud particle moments $\rho L_j$ as
\begin{equation}
  {\rm v}_{{\rm dr},j} = \xi\frac{\rhod}{\rho\,c_T}\frac{L_{j+1}}{L_j}
  \label{eq:vj}\ .
\end{equation}
The cloud particle moments are updated according
to the following explicit upwind advection scheme. We first calculate all
vertical moment fluxes $\phi_{j,i}=\phi_j(z_i)$ via Eq.\,(\ref{eq:jj})
and then apply 
\begin{eqnarray}
  \rho L_{j,\,i}^{(n)} &\!\!\!=\!\!\!& \rho L_{j,\,i}^{(n-1)}
  \hspace*{2mm}+ \frac{\Delta t}{\Delta z}
    \left(\phi_{j,\,i+1}^{\,(n-1)} - \phi_{j,\,i}^{\,(n-1)}\right) \\
  \rho {L^s_{3,\,i}}^{\!\!\!(n)} &\!\!\!=\!\!\!& \rho {L^s_{3,\,i}}^{\!\!\!(n-1)}
  + \frac{\Delta t}{\Delta z}
    \left({b^s_{i+1}}^{\!\!\!\!(n-1)}\phi_{3,\,i+1}^{\,(n-1)} 
         -{b^s_i}^{(n-1)} \phi_{3,\,i}^{\,(n-1)}\right) \ ,
\end{eqnarray}
where the notation $f^{\,(n)}_i$ is some quantity on grid point $i$ at
time $t^{\,n}$, $\Delta t\!=\!t^{\,n}-t^{\,n-1}$ the timestep and
$\Delta z$ the vertical extension of the considered atmospheric cell.

\section{Verification tests} 
\label{app:tests}
\vspace*{-1mm}
We have carefully checked our diffusion solver against analytical test
problems and by cross-checking the results from the explicit and
implicit solvers.  Figures \ref{testfig3} and \ref{testfig4} show two
test problems on domain $z\!=\![0,1]$, with constant $\nH\!=\!1$ and
$D\!=\!1$. The black thin lines are the overplotted analytic
solutions, showing excellent agreement. The tests use an equidistant
$z$-grid with 101 points, and can be computed within less than
1\,CPU-sec.

The convergence of the full cloud formation model was studied by
comparing the results obtained with different initial conditions (see
Sect.~\ref{sec:ic}). Figure~\ref{fig:init} shows the results of two
models for $T_{\rm eff}\!=\!1800K$, $\log\,g\!=\!3$,
${\beta^{\prime}}\!=\!1$ and solar abundances. The initially `full'
model shows a massive cloud formation event just after onset.  The
cloud formation is most effective at high densities and low
temperatures, causing transient minimums of $\ek(z)$.  The initially
`empty' model needs more time to start forming clouds because the
condensable elements first need to be transported upwards by
diffusion, resulting in a more gradual onset of cloud formation.  The
final states after $t\!=\!8000\,$days are identical in both cases,
where the $\ek(z)$ decrease monotonically with height and have zero
gradients at the upper boundary, as it should be in the
time-independent case, see Sect.~\ref{sec:tindep}.

\section{Diffusion coefficients in the literature}
\label{s:DG_A}


{Diffusion, in principle, is a microscopic process driven
  by particle concentration gradients $\nabla c_{\!j}\,$, where for
  example $c_{\rm CO}=n_{\rm CO}/n_{\rm tot}$ for $j\!=\!\rm CO$ and
  $n_{\rm tot}\!=\!\sum n_j$ is the total particle density.}
Such gradients can result from
gravity \citep{Zahnle2016}, from chemical processes
\citep{2000Icar..143..244M} and from cloud condensation as shown in
this paper.
Diffusion will always counteract these concentration gradients.
Experiments have been conducted to measure diffusion constants for
gases relevant for solar system planets. \citet{Lamb2011} provide, for
example, the {gas-kinetic diffusion coefficient for water
  molecules near sea level in the Earth atmosphere} ($D_{\rm
  H2O}\!\approx\!2\times10^{-1}\rm\,cm^2/s$) and note that it varies
inversely proportional with the atmospheric pressure.

{The effect of mixing on larger scales has been modelled
  differently in different communities, and terminology is usually not
  unique. Often, a quasi-diffusive approach is used where the
  diffusion constant is replaced by a function of height or 
  density/pressure.}  Transport of matter due to turbulent mixing has
been termed 'turbulent diffusion' in protoplanetary disk modelling,
describing the averaged effect of advection of the individual
turbulent eddies \citep[e.g.][]{2004ApJ...614..960S} and as 'eddy
diffusion' in planetary atmosphere modelling.

Studying solar system giants, \citet{2000Icar..143..244M} have
demonstrated that ISO observations of hydrocarbon molecules in
Saturn's atmosphere can be well fitted by assuming an eddy diffusion
coefficient of
\begin{equation}
  D_{\rm mix} = 1.838\times10^7{\rm\,cm^2/s}\,
       \left(\frac{7.213\times10^{11}{\rm\,cm^{-3}}}{n_{\rm tot}}\right)^\beta
\end{equation}
with slope $\beta$ between 0.3 and 0.7, i.e. $D_{\rm mix}$ {\it
  increases} with height.


\citet{2001ApJ...556..872A} consider an
equilibrium between upward mixing of vapour in the gas phase and
gravitational settling of particles condensed from the vapour. Their
Eq.\,(4) reads
\begin{equation}
    -D_{\rm mix}\,\frac{d}{dz} \left(q_c+q_v\right) 
    -f_{\rm sed}\,w_\star\,q_c = 0
  \label{eq:AMK}
\end{equation}
{where $q_v$ and $q_c$ are the mixing ratios of vapour and condensate,
respectively (moles of vapour/condensate per mole of atmosphere). 
$f_{\rm sed}\,w_\star$ represents an average sedimentation velocity of
the condensed particles with $f_{\rm sed}$ being adjusted as needed.
We note that Eq.\,(\ref{eq:AMK}) is very similar to our
Eq.\,(\ref{eq4}) for the case where cloud and gas particles are equally
affected by eddy-diffusion. Their eddy-diffusion coefficient 
$D_{\rm mix}$ is defined according to (\citealt{1985rapm.book..121G})
as
\begin{equation}
  D_{\rm mix} = \frac{H}{3} \left(\frac{\ell_{\rm mix}}{H}\right)^{4/3} 
    \left(\frac{RF_{\rm conv}}{\mu\,\rho\,c_p}\right)^{1/3}
\end{equation}
where $H\!=\!RT/(\mu\,g)$ is the pressure scale height,
$\ell_{\rm mix}$ the mixing length, $c_p\!=\!(f+2)R/(2\mu)$
is the isobaric specific heat, $f$ is the mean degree
of freedom of the gas particles, $R$ the universal gas constant, and
$F_{\rm conv}$ is the convective heat flux. \cite{2001ApJ...556..872A}
assume $F_{\rm conv}\!=\!\sigma T_{\rm eff}^4$, i.e.\ they assume that
the atmosphere is fully convective, which leaves open the problem of
what to do in radiative layers, for example in brown dwarf
atmospheres.  \citet{2018ApJ...854..172C} note that the factor 1/3 is
chosen to match observations from giant gas planets.  The mixing
length $l_{mix}$ is calculated as}
\begin{equation}
  l_{\rm mix} = H\cdot \max\{\Lambda, \Gamma/\Gamma_{\rm adb}\}
\end{equation}
in \citet{2001ApJ...556..872A}, where $\Gamma$ and $\Gamma_{adb}$ are
the local and dry adiabatic lapse rates, respectively, and $\Lambda$
is the minimum scaling applied to $\ell_{mix}$, chosen to be $0.1$.
The average sedimentation velocity is $w_\star\!=\!D_{\rm
  mix}/\ell_{\rm mix}$. We note that, if $H$, $c_P$ and $\ell_{\rm
  mix}$ are constants, the diffusion constant scales as $D_{\rm mix}
\propto n_{\rm tot}^{-1/3}$ in the Ackerman \& Marley models,
i.e.\ their $D_{\rm mix}$ increases with
height. \citet{2018ApJ...854..172C} use a similar approach, deriving
the convective heat flux from their simulations inside the convective
zone. In radiative layers, however, they assume $F_{\rm
  conv}\!=\!10^{-6}\,\sigma T_{\rm eff}^4$ to account for the effect
of convective overshooting. This approach enables them to model
secondary cloud layers.

\cite{Zahnle2016} use a combination of gas-kinetic diffusion and
eddy-diffusion, which is standard in 1D chemical models for
planetary atmospheres
\begin{eqnarray}
  n_{\rm tot} \frac{dc_i}{dt} &\!=\!&
    P_i-L_i \,-\frac{d}{dz} \phi_i \\
    \phi_i &\!=\!&
    \left(\frac{\mu g}{kT}-\frac{m_i g}{kT}\right) b_{ia}\,c_i
    ~-~ \big(b_{ia}+D_{\rm mix}n_{\rm tot}\big)\frac{d}{dz} c_i
  \label{Zahnle}  
\end{eqnarray}
{Here, $b_{ia}\!=\!D_{ia}/n_{\rm tot}\,\rm[cm^{-1}s^{-1}]$ is the
  binary diffusion coefficient and $D_{ia}\,\rm[cm^2s^{-1}]$ the
  gas-kinetic diffusion coefficient for particles of kind $i$ in a
  background atmosphere $a$ with mean molecular weight $\mu$. $P_i$
  and $L_i\,\rm[cm^{-3}s^{-1}]$ are the chemical production and loss
  rates and $\phi_i\,\rm[cm^{-2}s^{-1}]$ is the total diffusive flux
  of particles of kind $i$. It is straightforward to verify that, in
  the absence of chemical processes and eddy-diffusion, molecules of
  different kinds $i$ would eventually relax towards independent
  stratifications $n_i\!=\!n_{i,0} \exp(-z/H_i)$ with scale heights
  $H_i\!=\!kT/(m_i\,g)$, whereas the background atmosphere would
  follow $n_{\rm tot}\!=\!n_{\rm tot,0} \exp(-z/H)$ with
  $H\!=\!kT/(\mu\,g)$.  This effect could be described as
  ``gravitational de-mixing'', resulting from the action of the force
  of gravity on a mixture of gases when only gas-kinetic diffusion
  is active. On the contrary, eddy diffusion counteracts this
  tendency and tends to homogenise the concentrations. The critical
  level below which the atmosphere is well-mixed is called the {\it
    homopause} and follows from $D_{ia}(z)\!=\!D_{\rm mix}(z)$. In
  their models, $D_{\rm mix}$ is a free constant between
  $(10^5-10^{10})\rm\,cm^2/s$. We note that, when the first
  term in Eq.\,(\ref{Zahnle}) is neglected, this matches our approach
  (Eq.\,\ref{eq:epsDiff}) with $D_{\rm gas}=D_{\rm mix}+D_{\rm
    micro}$.  A similar description has been used by \citet[][see
    their Eq.\,23]{2016ApJS..224....9R}.}

{Using 2D radiative-hydrodynamics simulations for brown dwarf
  atmospheres, \citet[][see their section 4.3 and Figs.\,13 and
    \,14]{2010A&A...513A..19F} have estimated eddy-diffusion
  coefficients from root-mean-square gas velocities $\langle v\rangle$
  as found in their models. Results range from about $10^5\,\rm cm^2/s$
  to $10^9\,\rm cm^2/s$, depending on the details of the conversion
  formula applied, and are relatively constant through the atmosphere.}


{\citet[][see their Eq.\,22]{2013A&A...558A..91P} follow tracer
particles in their 3D GCM models for the hot Jupiter HD189733b to
provide approximate eddy-diffusion coefficients as function of gas
pressure as
\begin{equation}
  D_{\rm mix} = \frac{5\times 10^8\,{\rm cm^2/s}}
               {\big(p/{\rm 1\,bar}\big)^{1/2}}
\end{equation}
Using a different 3D GCM code with time-dependent cloud formation
theory for HD189733b, \cite{Lee2015} provide approximate
eddy-diffusion coefficients (their Fig.~3) fitted with a powerlaw
as $D_{\rm mix}\!\propto\!p^{-0.65}$, again showing increasing
eddy-diffusion coefficients with height.}

{\cite{2018ApJ...866....1Z} use 3D atmosphere models to study
tracer particles which have a certain (chemical) lifetime.  They suggest
$D_{\rm mix}\!\sim\!\tau_{\rm c}$ when the chemical lifetime $\tau_{\rm
  c}$ of a tracer species is short, and $D_{\rm mix}$ is constant.
Regime-dependent $D_{\rm mix}$ parameterisations are provided.}

{Other parameterisations are used in modelling planet-forming disks.
In \citet{2011A&A...534A..73Z}, the parameterised diffusion coefficient is
\begin{equation}
  D_{\rm mix} = \alpha\,c_T H_p
  \label{diffusion-coefficient-gas-zsom}
\end{equation}
where $\alpha$ is the dimensionless {viscosity parameter introduced by
\citet{Shakura1973}}
\begin{equation}
  \alpha = \frac{\langle v_z\rangle}{c_T} + \frac{H^2}{4 \pi \rho c_T^2}
  \ .
\end{equation}
If magnetic fields $H$ are neglected, then
Eq.\,\eqref{diffusion-coefficient-gas-zsom} reduces to our
Eq.\,\eqref{eq:defDmix} for the eddy-diffusion
coefficient. $\alpha\!\approx\!10^{-6}-10^{-2}$ is treated as an
adjustable parameter in disk simulations.  In \cite{Youdin2007}, the
gas diffusion coefficient takes the form
\begin{equation}
  D_{\rm mix} = \langle v_z\rangle^2 \tau_{\rm eddy}
\end{equation}
where $\tau_{\rm eddy}$ is the turbulent eddy turnover timescale.}

%

\section{Cloud Opacity Estimations}
\label{sec:opac}

As an order of magnitude estimate for cloud particle opacities, we
consider small spherical particles with optical constants for
astronomical silicates \citep{Draine1984,Laor1993}, using a MRN
\citep{Mathis1977} size distribution $f(a)\!\propto\!a^{-3.5}$ between
$a_{\rm min}\!=\!0.005\,\mu$m and $a_{\rm max}\!=\!0.25\,\mu$m, which
is a standard for the dust in the interstellar medium. Opacities are
calculated with Mie theory and listed in Table~\ref{MRN}.

\begin{table}[!h]
\caption{MRN astronomical silicate dust extinction opacities
  $\kappa_\nu^{\rm ext}$ for selected wavelengths $\lambda$, see
  text for references.}
\vspace*{-2mm}
\label{MRN}
\centering
\begin{tabular}{c|c}
$\lambda\rm\,[\mu m]$ & $\kappa_\nu^{\rm ext}\rm\,[cm^2/g(dust)]$\\
\hline
&\\[-2.2ex]  
0.55 & 16000 \\
1    & 3700 \\
5    & 230 \\
10   & 2700 \\
30   & 380 \\
100  & 30 
\end{tabular}
\vspace*{-2mm}
\end{table}

\noindent Cloud opacities in the atmospheres of brown dwarfs and
exoplanets will differ from those values because of deviations in
material composition, size and shape distribution. Typical opacity
values for larger particles in protoplanetary discs at
$\lambda\!=\!1\,\mu$m are expected to range from several 100 to
several $1000\,\rm cm^2/g(dust)$, see e.g.\ Fig.~3 in
\citet{Woitke2016}.

Given the total column densities of cloud particles found in our
models (see Table~\ref{tab:oldnew} and Fig.~\ref{fig:coldens}), we
conclude that the clouds in our models are increasingly optically
thick towards lower effective temperatures. We estimate that the
clouds become optically thick at $\lambda\!=\!550\,$nm for $T_{\rm
  eff}\!\la\!2500\,$K, at $\lambda\!=\!1\,\mu$m and $10\,\mu$m for
$T_{\rm eff}\!\la\!2000\,$K, but are considerably more transparent at
e.g.~$\lambda\!=\!5\,\mu$m and beyond $\lambda\!\ga\!30\,\mu$m.

\end{appendix}

\end{document}

%% file: farben.txt
\definecolor{GreenYellow}  {cmyk}{0.15,0,0.69,0}
\definecolor{Yellow}{cmyk}{0,0,1,0}
\definecolor{Goldenrod}{cmyk}{0,0.10,0.84,0}
\definecolor{Dandelion}{cmyk}{0,0.29,0.84,0}
\definecolor{Apricot}  {cmyk}{0,0.32,0.52,0}
\definecolor{Peach}    {cmyk}{0,0.50,0.70,0}
\definecolor{Melon}    {cmyk}{0,0.46,0.50,0}
\definecolor{YellowOrange}  {cmyk}{0,0.42,1,0}
\definecolor{Orange}   {cmyk}{0,0.61,0.87,0}
\definecolor{BurntOrange}   {cmyk}{0,0.51,1,0}
\definecolor{Bittersweet}   {cmyk}{0,0.75,1,0.24}
\definecolor{RedOrange}{cmyk}{0,0.77,0.87,0}
\definecolor{Mahogany} {cmyk}{0,0.85,0.87,0.35}
\definecolor{Maroon}   {cmyk}{0,0.87,0.68,0.32}
\definecolor{BrickRed} {cmyk}{0,0.89,0.94,0.28}
\definecolor{Red} {cmyk}{0,1,1,0}
\definecolor{OrangeRed}{cmyk}{0,1,0.50,0}
\definecolor{RubineRed}{cmyk}{0,1,0.13,0}
\definecolor{WildStrawberry}{cmyk}{0,0.96,0.39,0}
\definecolor{Salmon}   {cmyk}{0,0.53,0.38,0}
\definecolor{CarnationPink} {cmyk}{0,0.63,0,0}
\definecolor{Magenta}  {cmyk}{0,1,0,0}
\definecolor{VioletRed}{cmyk}{0,0.81,0,0}
\definecolor{Rhodamine}{cmyk}{0,0.82,0,0}
\definecolor{Mulberry} {cmyk}{0.34,0.90,0,0.02}
\definecolor{RedViolet}{cmyk}{0.07,0.90,0,0.34}
\definecolor{Fuchsia}{cmyk}{0.47,0.91,0,0.08}
\definecolor{Lavender} {cmyk}{0,0.48,0,0}
\definecolor{Thistle}{cmyk}{0.12,0.59,0,0}
\definecolor{Orchid}{cmyk}{0.32,0.64,0,0}
\definecolor{DarkOrchid}{cmyk}{0.40,0.80,0.20,0}
\definecolor{Purple}{cmyk}{0.45,0.86,0,0}
\definecolor{Plum}{cmyk}{0.50,1,0,0}
\definecolor{Violet} {cmyk}{0.79,0.88,0,0}
\definecolor{RoyalPurple} {cmyk}{0.75,0.90,0,0}
\definecolor{BlueViolet}{cmyk}{0.86,0.91,0,0.04}
\definecolor{Periwinkle}{cmyk}{0.57,0.55,0,0}
\definecolor{CadetBlue}{cmyk}{0.62,0.57,0.23,0}
\definecolor{CornflowerBlue}{cmyk}{0.65,0.13,0,0}
\definecolor{MidnightBlue}{cmyk}{0.98,0.13,0,0.43}
\definecolor{NavyBlue} {cmyk}{0.94,0.54,0,0}
\definecolor{RoyalBlue}{cmyk}{1,0.50,0,0}
\definecolor{Blue}{cmyk}{1,1,0,0}
\definecolor{Cerulean} {cmyk}{0.94,0.11,0,0}
\definecolor{Cyan}{cmyk}{1,0,0,0}
\definecolor{ProcessBlue} {cmyk}{0.96,0,0,0}
\definecolor{SkyBlue}{cmyk}{0.62,0,0.12,0}
\definecolor{Turquoise}{cmyk}{0.85,0,0.20,0}
\definecolor{TealBlue} {cmyk}{0.86,0,0.34,0.02}
\definecolor{Aquamarine}{cmyk}{0.82,0,0.30,0}
\definecolor{BlueGreen}{cmyk}{0.85,0,0.33,0}
\definecolor{Emerald}{cmyk}{1,0,0.50,0}
\definecolor{JungleGreen} {cmyk}{0.99,0,0.52,0}
\definecolor{SeaGreen} {cmyk}{0.69,0,0.50,0}
\definecolor{Green}{cmyk}{1,0,1,0}
\definecolor{ForestGreen} {cmyk}{0.91,0,0.88,0.12}
\definecolor{PineGreen}{cmyk}{0.92,0,0.59,0.25}
\definecolor{LimeGreen}{cmyk}{0.50,0,1,0}
\definecolor{YellowGreen} {cmyk}{0.44,0,0.74,0}
\definecolor{SpringGreen} {cmyk}{0.26,0,0.76,0}
\definecolor{OliveGreen}{cmyk}{0.64,0,0.95,0.40}
\definecolor{RawSienna}{cmyk}{0,0.72,1,0.45}
\definecolor{Sepia}{cmyk}{0,0.83,1,0.70}
\definecolor{Brown}{cmyk}{0,0.81,1,0.60}
\definecolor{Tan} {cmyk}{0.14,0.42,0.56,0}
\definecolor{Gray}{cmyk}{0,0,0,0.50}
\definecolor{Black}{cmyk}{0,0,0,1}
\definecolor{White}{cmyk}{0,0,0,0}

\definecolor{StaubBraun}{cmyk}{0,0.61,0.87,0.20}
\definecolor{lyellow}{cmyk}{0,0,0.2,0}
\definecolor{lred}{cmyk}{0,0.4,0.4,0}
\definecolor{lgray}{cmyk}{0,0,0,0.15}
\definecolor{lgreen}{cmyk}{0.3,0,0.8,0}
\definecolor{lblue}{cmyk}{0.4,0,0,0}
\definecolor{dgreen}{rgb}{0.05,0.5,0.1}
\definecolor{dred}{rgb}{0.85,0.0,0.0}
\definecolor{dblue}{rgb}{0.0,0.0,0.6}
\definecolor{brown}{rgb}{0.6,0.1,0.1}